\documentclass[12pt]{article}
\usepackage{scicite}
\usepackage{times}

\newenvironment{sciabstract}{%
\begin{quote} \bf}
{\end{quote}}

\newcounter{lastnote}

\usepackage{graphicx}
\graphicspath{{./EPJ-Figs/}}
\DeclareGraphicsExtensions{.pdf,.jpeg,.png}
\usepackage{epstopdf}
\usepackage{subfig}
\usepackage{url}
\usepackage{floatrow}
\usepackage{amsmath}
\usepackage[usenames,dvipsnames]{color}
\usepackage{bigfoot}
\usepackage[english]{babel}
\usepackage[utf8x]{inputenc}
\usepackage[colorinlistoftodos]{todonotes}

\newcommand{\beginsupplement}{
        \setcounter{table}{0}
        \renewcommand{\thetable}{S\arabic{table}}
        \setcounter{figure}{0}
        \renewcommand{\thefigure}{S\arabic{figure}}}

\title{Segregated interactions in urban and online space}

\author
{Xiaowen Dong,$^{1,2\dagger}$ Alfredo J. Morales,$^{2,3\dagger}$ Eaman Jahani,$^{4\dagger}$\\ Esteban Moro,$^{2,5}$ Bruno Lepri,$^{6}$ Burcin Bozkaya,$^{7,8}$\\ Carlos Sarraute,$^{9}$ Yaneer Bar-Yam,$^{3}$ Alex `Sandy' Pentland$^{2}$\\
\\
\normalsize{$^{1}$Department of Engineering Science, University of Oxford, Oxford, UK}\\
\normalsize{$^{2}$MIT Media Lab, Cambridge, MA, USA}\\
\normalsize{$^{3}$New England Complex Systems Institute, Cambridge, MA, USA}\\
\normalsize{$^{4}$MIT Institute for Data, Systems and Society, Cambridge, MA, USA}\\
\normalsize{$^{5}$Department of Mathematics and GISC, Universidad Carlos III de Madrid, Legan\'es, Spain}\\
\normalsize{$^{6}$Mobile and Social Computing Lab, Fondazione Bruno Kessler, Trento, Italy}\\
\normalsize{$^{7}$Behavioral Analytics \& Visualization Lab, Sabanc{\i} University, Istanbul, Turkey}\\
\normalsize{$^{8}$New College of Florida, Sarasota, FL, USA}\\
\normalsize{$^{9}$Grandata Labs, Buenos Aires, Argentina}\\
\\
\normalsize{$^\dagger$These authors contributed equally to this work.}
}

\date{}

\begin{document}

\baselineskip24pt

\maketitle 

\begin{sciabstract}
Urban income segregation is a widespread phenomenon that challenges societies across the globe. Classical studies on segregation have largely focused on the geographic distribution of residential neighborhoods rather than on patterns of social behaviors and interactions. In this study, we analyze segregation in economic and social interactions by observing credit card transactions and Twitter mentions among thousands of individuals in three culturally different metropolitan areas. We show that segregated interaction is amplified relative to the expected effects of geographic segregation in terms of both purchase activity and online communication. Furthermore, {we find that segregation increases with difference in socio-economic status but is asymmetric for purchase activity, i.e., the amount of interaction from poorer to wealthier neighborhoods is larger than vice versa.} Our results provide novel insights into the understanding of behavioral segregation in human interactions with significant socio-political and economic implications.
\end{sciabstract}

\section*{Introduction}
Residential segregation has historically been associated with societal issues such as economic, educational, and health inequalities \cite{segregationeffect,Sampson2012}; as a consequence, it has been a central focus in social, economic and political sciences \cite{Burgess28,Duncan55,Schelling71}.
Recent studies show that while racial segregation seems to be decreasing in the United States \cite{Glaeser12,Firebaugh16}, income inequality has been simultaneously rising \cite{Reardon2011,Florida2016}. According to the Stanford Center on Poverty and Inequality, 1\% of the American population held 21\% of all the income in 2012, which is more than double of what they held in 1970 (8.4\%). This change is coupled with a sharp increase in residential segregation by income \cite{Fry2012}. In forty years, the number of American families living in middle-income neighborhoods went from 65\% down to 43\% in large metropolitan areas. Families are thus increasingly living in either extremely poor or rich neighborhoods, endangering the existence and stability of the middle classes \cite{incomesegregation}.

In order to quantify residential segregation, American census reports \cite{Iceland02} calculate twenty different indexes across five dimensions, namely: \textit{evenness}, \textit{exposure}, \textit{concentration}, \textit{centralization} and \textit{clustering} \cite{Massey98}. These metrics are mostly based on static census data and do not reflect patterns in an activity or behavioral space. At the same time, regardless of whether they involve physical space or not, restrictions on any type of social interaction may be considered as forms of segregation \cite{Freeman78}.
These, together with the increasing availability of data sources resulting from human activities \cite{Lazer09,Pentland14}, have led to an increasing number of studies on modern forms of segregation in spaces beyond residential neighborhoods.
Most notably, recent works have shown that there exists clear separation between different ethnic or income groups in everyday activities such as visitation of urban areas \cite{Wong11,Bora14,Boterman16,Wang16,Yip16,Wang18,Morales19} or consumption of online information \cite{Gentzkow11,Morales15,Bakshy15,Quattrociocchi16a,Bastos17,Bail18}, leading to the so-called ``echo chambers'' or ``filter bubbles'' \cite{Levy19}.

While the literature mainly focuses on the limited exposure of certain socio-demographic and wealth groups to the others, the restriction on interactions between these groups \cite{Blumenstock13} remains rather unexplored, possibly due to the lack of large-scale interaction data. 
{The recent work of Morales \emph{et al.} \cite{Morales19} has shown that groups of different income levels have differentiated topics of conversation, and that exposure limited by segregated interactions both offline and online is a key variable for homogeneization. Along a similar line of investigation,} 
{and following recent studies using similar data sources in analysing urban mobility and behavior \cite{Leo16,Clemente18,Dong18,Luo16},}
we combine in the present paper large-scale credit card transaction and Twitter data sets to study income segregation in daily purchase activities and online communication, thus capturing two explicit interactions in economic and social behavior.
We analyze how the patterns of segregation in both offline and online activities are intertwined, and vary with respect to both difference in socio-economic status and geographical distance.
We demonstrate the consistency in these patterns by examining different cultural and political contexts, in three large metropolitan areas from Europe, Latin America, and North America\footnote{Credit card transaction data are only available for the European and Latin American cities.}.
Although we do not have a direct matching of individuals between the transaction and Twitter data sets, we study behaviors at the collective scale by aggregating the data by urban administrative neighborhoods, for which socio-economic status can be obtained from national census data.

The main contributions of the present paper are three-fold. First, we show that segregation in behavioral interactions is amplified with respect to the expected effect of geographic segregation, in terms of both purchase activity and online communication. Second, we analyse segregation with respect to socio-economic status and geographical distance, where we found that segregation is most pronounced between extreme income groups. {Finally, we demonstrate that segregation is asymmetric for purchase activity, where the amount of interactions from poorer to wealthier neighborhoods is larger than the other way around.}
These findings provide a new angle to study modern forms of segregated behavior, with implications on urban planning, policy-making, and inequality reduction.

\section*{Results}
\begin{figure}[t]
      \centering
      {\includegraphics[width=16cm]{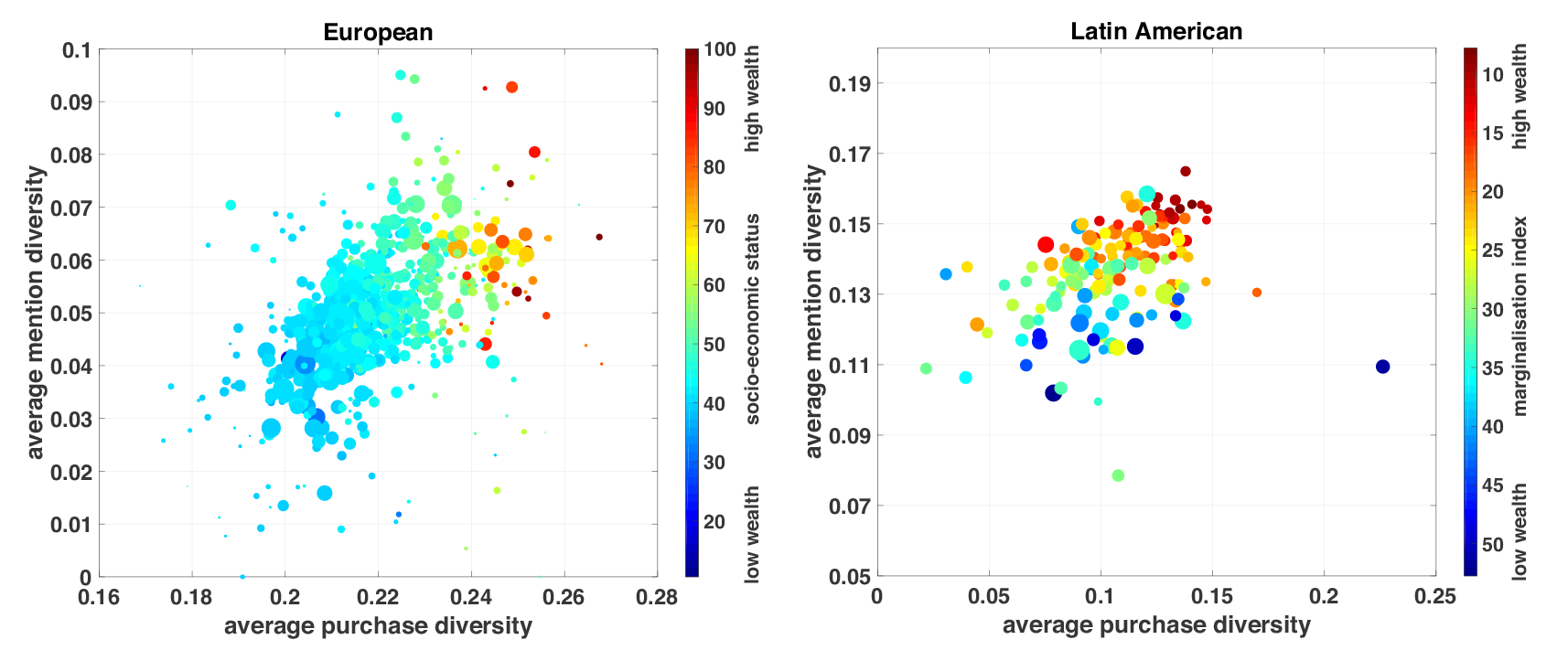}}
        \caption{Neighborhood-level average diversity of purchase{s} and Twitter mention{s} in the European and Latin American metropolitan areas. The size of the dots is proportional to the census population of the neighborhood and the color code indicates its wealth level. The correlation between both types of diversity is 0.45 and 0.38 for the European and Latin American case, respectively.}
        \label{fig:fig1}
\end{figure}

Human patterns of exploration in urban and social spaces are linked to both individual and regional economic growth \cite{Eagle10b,Singh15,Dong16}. We measure exploration by means of the diversity of purchases and Twitter communication via mentions. In order to measure diversity, we first characterize each individual with a pair of vectors whose elements represent either shops (in the case of purchases) or other individuals (in the case of Twitter mentions), and count the number of times individuals purchase at each shop or communicate with other individuals, respectively. We then measure, as explained in Materials and Methods, the individual diversity as the Shannon entropy of each vector. Figure~\ref{fig:fig1} shows a scatter plot with the aggregate neighborhood diversity of purchases and Twitter mentions after averaging over the individuals who live in each neighborhood of the European and Latin American cities. 
The average neighborhood diversity of both types of behaviors is positively correlated with each other ($r=0.45$ in Europe and $r=0.38$ in Latin America), as well as with the neighborhood socio-economic status\footnote{{In the European city, the correlations between purchases and Twitter mentions' diversity and socio-economic status are $r=0.72$ and $r=0.38$, respectively; In the Latin American city, the correlations between purchases and Twitter mentions' diversity and marginalization index (negative socio-economic status) are $r=-0.70$ and $r=-0.33$, respectively; In the Northern American city, the correlation between Twitter mentions' diversity and median household income (approximation of socio-economic status) is $r=0.40$.}}. This indicates that people living in poorer neighborhoods are less exploratory in their purchase and online activities, suggesting that they live in physically and virtually confined spaces.

\begin{figure}[t]
      \centering
      {\includegraphics[width=16cm]{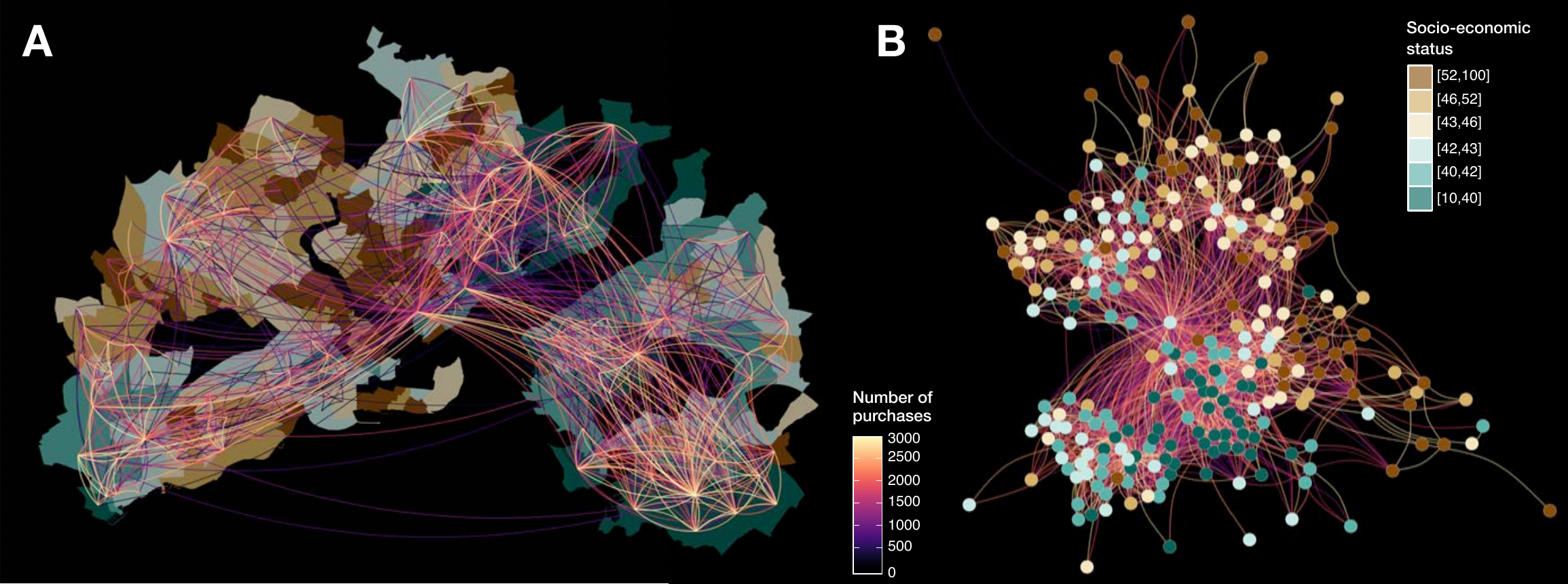}}
      \caption{Purchase interactions among some of the neighborhoods in the European metropolitan area. Each neighborhood is shown as an area on the map (panel A) and as a node in a graph (panel B), and it is in both cases color-coded by its socio-economic status. Each curve in both panels represents an interaction between a pair of neighborhoods, and it is color-coded by the number of purchases made by residents of one neighborhood in another. The directions of the curves are represented by their convexity: curves of convex shape represent interactions from the left end point to the right end point, while those of concave shape represent interactions from right to left.}
      \label{fig:fig2}
\end{figure}

The confinement of physical and virtual spaces is associated with segregation by income. We analyze this relationship by creating networks of interactions among neighborhoods based on purchases and Twitter mentions. In both networks, nodes represent neighborhoods whose socio-economic status is obtained from census data. Edges represent either the number of purchases made by customers living in neighborhood $i$ at stores in neighborhood $j$, or the number of tweets directed from users living in $i$ to users living in $j$. {To account for potential bias in the sampling of the users in the data sets, we use population-weighted versions of the interaction networks (see Materials and Methods for data statistics and construction of population-weighted interaction networks).} 
Figure~\ref{fig:fig2} displays an illustration of the purchase network for some of the neighborhoods in the European metropolitan area.
By analyzing the structure of these networks and the distribution of edges among neighborhoods, we are able to observe patterns of urban mixing or segregation.

In order to quantify segregation, {we put the neighborhoods into ten groups according to their socioeconomic status, where all groups have an equal number of neighborhoods. We then create mixing matrices whose elements show the aggregate number of interactions between the ten groups \cite{Newman03}.}
We further normalize the mixing matrix for both behaviors into a stochastic matrix whose elements show the probability of directed interaction among pairs of socio-economic groups (see Materials and Methods and Appendix for the construction and visualization of the mixing matrices).
We found that most of the interactions occur within groups of the same socio-economic status. We quantify such preference by calculating the assortativity coefficient of the mixing matrices \cite{Newman03}. 
A coefficient of 1 indicates a perfectly assortative network while 0 indicates random mixing patterns\footnote{A perfectly disassortative network, in which every edge connects two vertices of different types, has a negative coefficient generally in the range between -1 and 0 \cite{Newman03}.}.
The intuition is that assortative matrices are dominated by entries along and close to the matrix diagonal, indicating a stronger preference for neighborhoods to interact with similar ones (hence segregation). For example, in the European metropolitan area the assortativity coefficient of the mixing matrices for purchase and Twitter mentions is 0.42 and 0.41, respectively, indicating a certain degree of segregation {(see Figure~\ref{fig:figS7}, Figure~\ref{fig:figS8}, and Figure~\ref{fig:figS9} for the mixing matrices in the European, Latin American, and Northern American cases, respectively).}

While the assortativity coefficient shows a global description of the network, it misses heterogeneity in its structure, such as differences in the amount of segregation among certain socio-economic groups. In order to capture such heterogeneity, we analyze the segregation between the highest and lowest socio-economic strata and progressively include the remaining socio-economic groups in both directions until reaching the whole network. {At each step, we consider a certain number of groups both at the top and bottom of the socio-economic distribution (hence focusing on a percentage of the neighborhoods) and measure their segregation with the assortativity coefficient (see Figure~\ref{fig:figS10}, Figure~\ref{fig:figS11}, and Figure~\ref{fig:figS12} in Appendix).} 
Figure~\ref{fig:fig3} (top left) shows the assortativity coefficient as a function of the percentage of neighborhoods considered at each extreme of the distribution, for both purchases and mentions' networks, in the European city. For comparison, we also include in the figure the expected results from two artificial networks generated by (i) simulating neighborhood-to-neighborhood interactions with a gravity-based model \cite{Krings09} and (ii) randomly reshuffling neighborhoods' socio-economic status in a null model (see Appendix for details on the construction of artificial interaction networks using gravity-based model and null model). {Similar to the empirical interaction networks, both artificial networks have been re-weighted according to neighborhood population.}

\begin{figure}[t]
      \centering
      {\includegraphics[width=16cm]{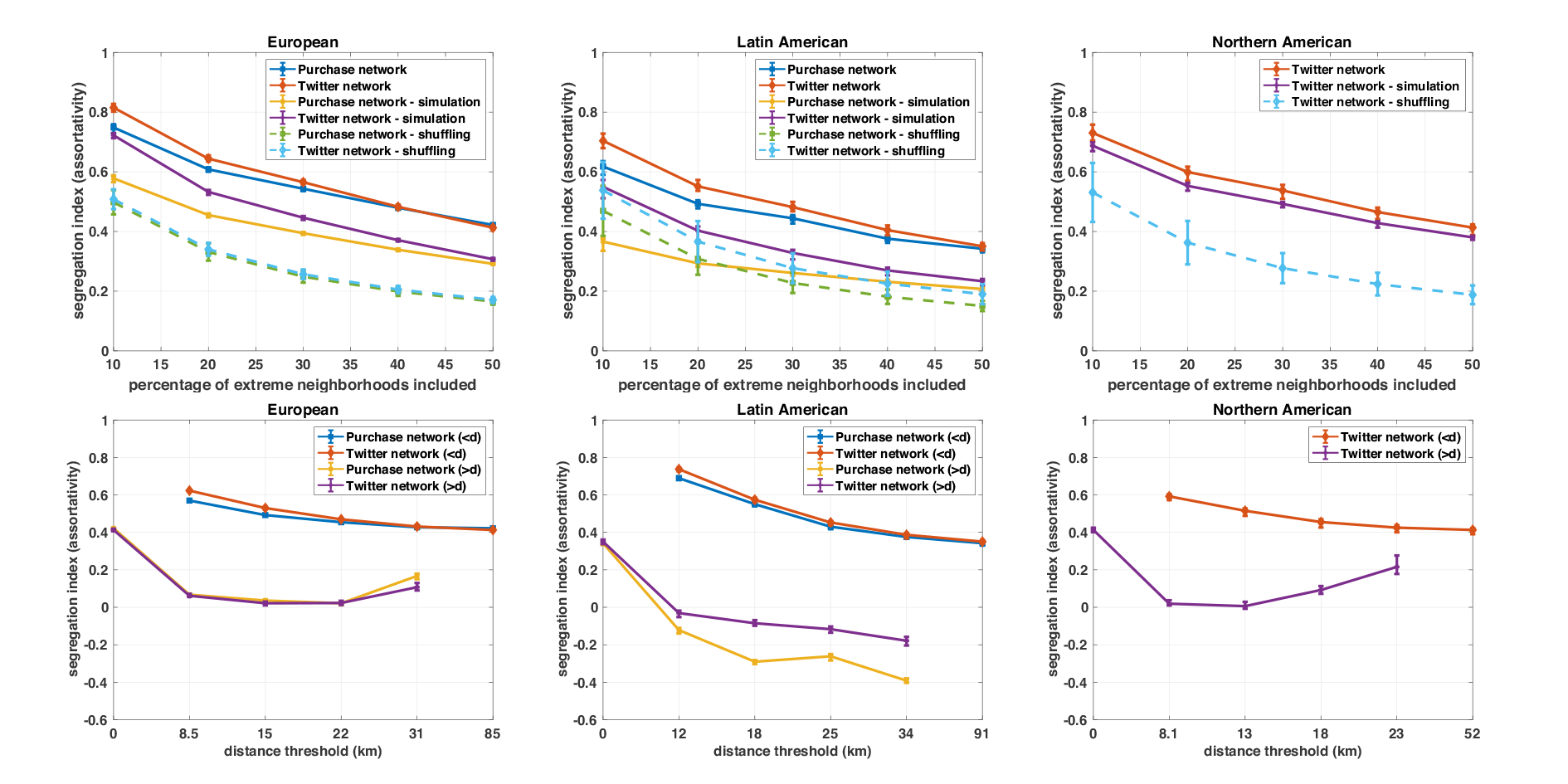}}
        \caption{(Top) The assortativity for different networks, as a function of {the} percentage of neighborhoods {with extreme socio-economic status} included in computation. (Bottom) The assortativity for different networks, as a function of {the} distance thresholds used for pruning edges in the interaction networks. {The error bars in the green and cyan curves correspond to the standard deviations, and all other error bars correspond to the 95\% confidence interval using a jackknife resampling technique as in \cite{Efron79}.}}
        \label{fig:fig3}
\end{figure}

The following observations can be made from Figure~\ref{fig:fig3} (top left). First, segregation is most pronounced between the highest and lowest socio-economic groups, which barely interact with each other, and decreases by including middle-class neighborhoods, which serve as ``social bridges'' between the richest and poorest parts of the society. Second, {segregation in interactions (blue and orange) is stronger than the one due to geographical distance (yellow and purple),} implying that it cannot be simply attributed to the segregated distribution of residential households in the city. Third, this segregation is also stronger than the one produced by the null model (green and cyan) showing that the patterns we observe are significant and not an artifact of the data. While the segregation in purchase patterns could be expected partially given the limitations that prices impose on people, the fact that they also tend to self-segregate on the Internet is interesting. {Furthermore, segregation online appears to be even stronger than offline especially between the highest and lowest socio-economic groups.} Similar patterns are also observed for the Latin and North American cities in Figure~\ref{fig:fig3} (top middle) and in Figure~\ref{fig:fig3} (top right).

{Geography and the organization of the physical urban space has been linked with segregation and inequality \cite{Toth19}. In order to further analyze the role of geographical distance in the segregation in interactions,} we measure the assortativity coefficient at multiple distances, by only considering subsets of neighborhood pairs that are either within a certain distance of $d$ km or beyond this distance, representing short- and long-distance interactions, respectively (see Appendix for details on the analysis procedure). 
The bottom row of Figure~\ref{fig:fig3} depicts the assortativity coefficients as a function of $d$ for the empirical networks of the three cities. It can be seen that both networks are predominately segregated due to short-distance interactions (blue and orange), with the network resulting from short-distance interactions having a consistently higher assortativity coefficient than that resulting from long-distance interactions (yellow and purple).
In the case of purchases, short-distance interactions are less costly in terms of time and money, and are dominated by daily activities such as groceries or banking. Interestingly, the same pattern holds for online behaviors, which could be dominated by the interaction of local social groups and is consistent with the finding in \cite{Bastos17}. Moreover, the positive assortativity score for long-distance interactions in the European and North American cases suggests that self-segregation might even exist between neighborhoods that reside further away.

\begin{figure}[t]
      \centering
      {\includegraphics[width=16cm]{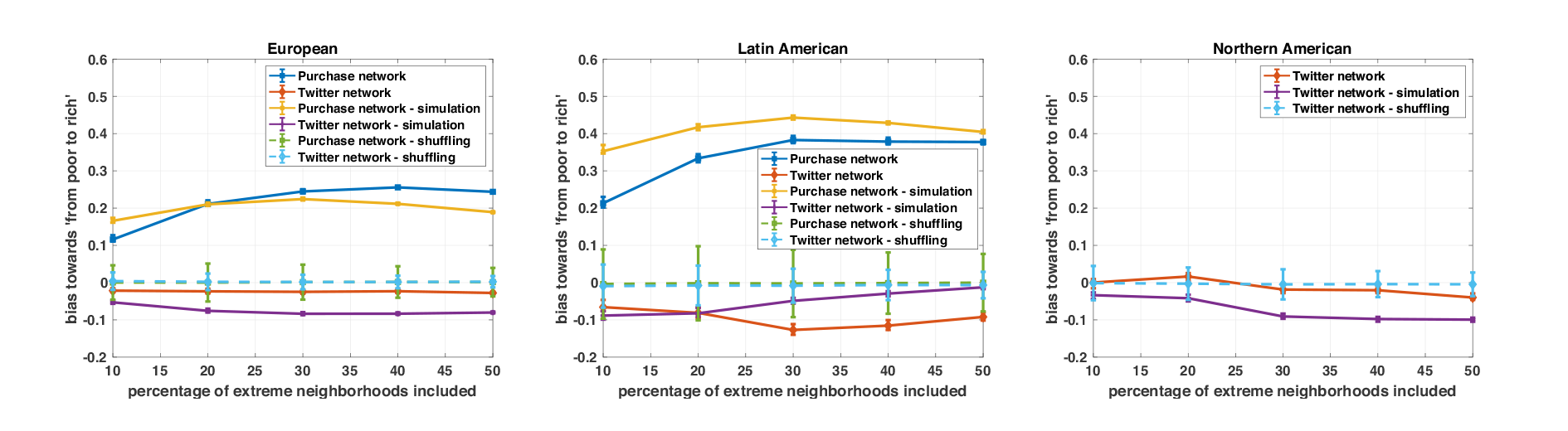}}
        \caption{Asymmetry in {the} interaction pattern{s} between the relatively poor and the relatively rich {segments of the population}. {The error bars correspond to the 95\% confidence interval using a jackknife resampling technique as in \cite{Efron79}.}}
        \label{fig:fig4}
\end{figure}

Apart from being segregated, 
{it is interesting to investigate whether interactions are symmetric among different socio-economic groups. To this end,} we measure the excess of interactions directed from poorer to richer areas, relative to the number of interactions in the opposite direction. This is quantified as the difference between the sums of the lower and upper triangles of the mixing matrices. Figure~\ref{fig:fig4} shows the poor-to-rich interaction bias as a function of the percentage of neighborhoods considered, following the same methodology presented in Figure~\ref{fig:fig3}, where we initially consider only the highest and lowest extremes of the wealth distribution and progressively include neighborhoods towards the middle. {The bias is positive in the case of offline purchases  (blue curve)}, and generally increases as we include the middle-low and middle-high wealth groups. 
Moreover, although still presenting asymmetric interactions, the less bias from the lowest to the highest wealth group, together with the most pronounced segregation between these two groups as shown in Figure~\ref{fig:fig3}, indicates a polarization of behavior in this case.
{On the other hand, online communication does not seem to exhibit such bias (orange curve)}.

We further investigate the robustness of the asymmetric relationship, by repeating the analysis on the two artificial networks introduced before: (i) simulated interaction networks based on a gravity model, and (ii) the ones produced by randomly reshuffling the neighborhoods' socio-economic status in a null model. 
{While in the former case the asymmetry persists for offline purchases (yellow curve in Figure~\ref{fig:fig4}), it disappears in the latter where the bias drops to zero (green curve).}
{For the European metropolitan area, when all neighborhoods are considered, the observed asymmetric pattern in offline behavior is more pronounced that the one produced by the gravity-based model.} This implies that the stronger tendency of interactions from the relatively poor to the rich cannot be simply attributed to a geographical factor. The same argument, however, does not hold for the case of the Latin American city, which is likely due to the observation that richer neighborhoods account for more stores in that case. 
{Although the observed asymmetric relationship for offline purchases might be influenced by a larger number of stores (and popular ones) in richer neighborhoods, it nevertheless suggests that at macro scale there seems to exist a hierarchy that is embedded in the behavioral interactions between different segments of the society.}

\section*{Discussion}
In summary, our results suggest that segregated patterns exist in both urban and online interactions between different socio-economic groups, and they seem stronger than that expected from merely the geographic distribution of residential households. 
{While residential neighborhoods in a region might consist of different socio-economic groups, interactions, both physically and socially, seem to take place more often between neighborhoods whose economic conditions are similar.} 
Such emerging behavior might be expected from purchase activities partially due to the constraints imposed by prices, but less so from online communication where boundaries are more likely self-imposed.

Indeed, while purchase behaviors are constrained by mobility, time and monetary resources associated with the spatial segregation patterns observed in our data, it was expected that online behavior would mitigate those constrains by creating a virtual {\em third-place} \cite{onlinecomm} where more diverse interaction would be possible. However, our results reinforce the recent findings that ``echo chambers'' in virtual space recreate and amplify the observed residential segregation in physical space. Nevertheless, it is still possible that segregation can be mitigated by encouraging virtual conversations and physical interactions between different groups. The promotion of such interactions might be critical in reducing segregation and prove more effective than simply an increase in the exposure to opposing views \cite{Bail18}.

More interestingly, we observe that the restrictions on interactions, in both urban and online space, are most pronounced among the extremes of the wealth distribution, but fuzzy for the middle classes, which might act as social bridges \cite{Dong18} distributing information across the social system.
Interactions across different segments of the society might therefore be promoted especially through the agency of groups with middle socio-economic status given their bridging roles.
{Furthermore, an asymmetric pattern of interaction for offline purchases seems to suggest the existence of a hierarchy at macro scale, where richer areas attract a disproportionately large amount of capical (see ``Segregation and economic inequality'' in Appendix).} This, in turn, is crucial for the creation of new economic opportunities. 
{As observed in Figure~\ref{fig:figS16}, for the European metropolitan area, a stronger segregation pattern in purchase interactions is linked with a higher level of inequality between neighborhoods in terms of their sales' revenues.} 
{This observation is worth of further investigation, with the possible implication that urban planners may consider a better strategy in allocating store locations for a more even distribution of capital,} which can be achieved by promoting tax segmentation.

Our analysis has limitations. Even though the number of users in the two data sets are correlated (see Figure~\ref{fig:figS1} and Figure~\ref{fig:figS2}), penetration rates of credit card and Twitter usage differ in neighborhoods of different socio-economic status. Richer neighborhoods 
tend to account for more samples in our data sets, an observation that is most pronounced for the Latin American city (see Figure~\ref{fig:figS2}). 
This leads to under-representation of population in neighborhoods with lower socio-economic status. 
{In this work, we have used population-weighted interaction networks to account for such sampling bias, and further investigation would be needed to fully assess its impact.}
Furthermore, {due to the culture and constraints of the different countries,} the credit card transaction data may only represent a fraction of the daily spending as people may choose to pay by cash in certain situations. 
{Finally, our socio-economic status data are obtained at the neighborhood level, and may not necessarily reflect the economic situation of individuals in the data sets.
Nevertheless, the general consistency between the results in three cities from three different continents across a period of several months suggests the validity of our findings in the contexts examined in the present study.}

\section*{Materials and Methods}

\subsection*{Data sets and pre-processing}
The credit card transaction data sets are provided by two major financial institutions, one in an European country and one in a Latin American country. Each record in the data set corresponds to one credit card transaction along with customer and store IDs, as well as the time (day, hour and minute) of the transaction and the spending amount in local currency. Additional information about the customers and stores are also made available, including customers' home location as well as store location and category.
{The customer-level data are pseudonymized such that each customer is represented by a pseudo-unique number, in a way similar to the pseudonymization of mobile phone call detail records (CDRs). In addition, all personally identifiable data attributes were removed before the data sets were provided to us.}

We focus our analysis on two large metropolitan areas of the two countries. As pre-processing steps, we first filter out foreign and online transactions to focus on local and physical activities. We then consider customers who made at least ten transactions in the data set. {In the European case, this leads to a set of 2.4 million records of individual credit card purchases from April to June 2013, made by 85 thousand individuals at 54 thousand stores. In the Latin American case, this consists of a set of 3.5 million records of individual credit card purchases from April to July 2013, made by 200 thousand individuals at 55 thousand stores.}

{We collect geo-localized Twitter data sets using Twitter's Streaming API \cite{twitterapi} from August 2013 to August 2014.} In the European case, it contains 76 million geo-localized tweets within the metropolitan area of interest, from 1.4 million Twitter users. In the Latin American case, it contains 10.3 million geo-localized tweets within the metropolitan area under study, from 422 thousand Twitter users. Finally, in the Northern American metropolitan area, it consists of 22.4 million geo-localized tweets, from 862 thousand Twitter users.

On Twitter, a user A can mention or reply to another user B in his post in which case the post contains B's username. This allows us to build neighborhood-level Twitter mention networks. For this purpose, {for the European case}, we select a subset of 20.1 million tweets containing user mentions or replies that are posted by 1 million users, for whom we are also able to infer their home locations. {For the Latin American and Northern American metropolitan areas, we collect 3.8 million tweets and 8.1 million tweets containing user mentions or replies that are posted by 260 thousand and 440 thousand users, respectively.}

{It is worth noting that in this paper we study interactions among urban neighborhoods of different socio-economic status. 
In all the three metropolitan areas we consider, neighborhoods are administrative districts of similar size used for census purposes. We have around 660 such neighborhoods in the European case, around 160 in the Latin American case, and around 190 in the Northern American case. 
{For the European city, the neighborhood-level socio-economic status is provided by a national institute in 2011,} which is a composite measure between 0 and 100 that quantifies the relative prosperity of the neighborhood based on a number of indicators such as income and education level. The higher the index, the more prosperous the neighborhood is. {In the Latin American case, we use the neighborhood-level marginalization index provided by a national institute in 2012 as an approximation of the (negative) socio-economic status,} namely, the higher the marginalization index, the lower the socio-economic status. {Finally, for the Northern American city, we use median household income provided by a national survey for the period of 2010-2014 to approximate the socio-economic status of the neighborhoods.}

Even though we do not have a matching between the individuals in the two different data sets, we are able to study both offline (purchases) and online (Twitter mentions) behavior at the level of administrative neighborhoods within the city. {Specifically, for the credit card data set, we associate the customers and stores with the neighborhoods in which they reside and are located, respectively. For the Twitter data set, the procedure for assigning a home neighborhood to a Twitter user is as follows. First, we map tweets to the neighborhoods. We do this by observing in which polygons (neighborhoods) the coordinates of the tweets of the user fall into. Second, we observe the times of the day in which the user tweets from each neighborhood. Third, we select the neighborhood that is used the most during night hours, i.e., from 8pm to 6am, as the home neighborhood of the user.} We then compute diversity scores and construct neighborhood-level interaction networks as described below.}

\begin{figure}
      \centering
      {\includegraphics[width=16cm]{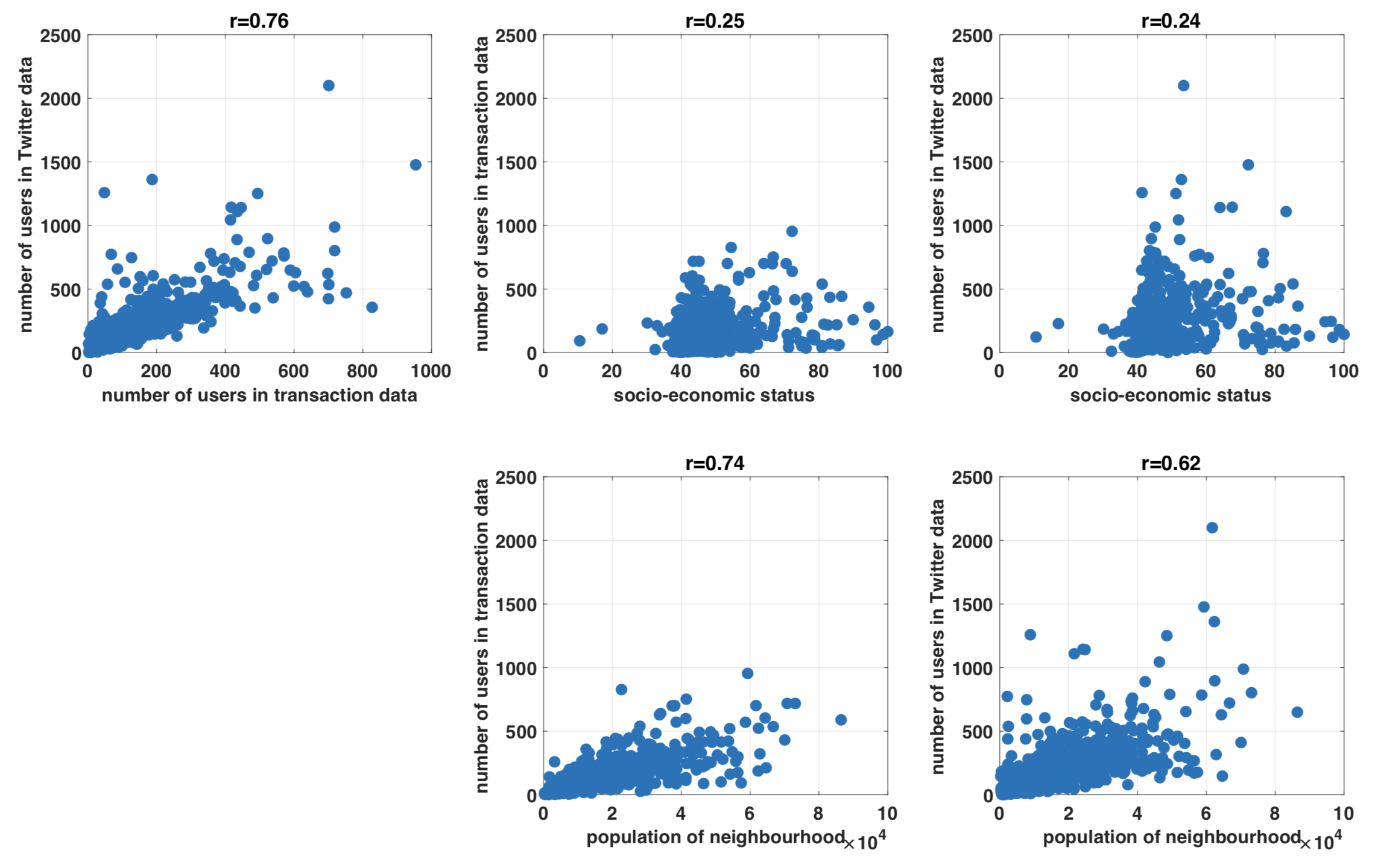}}
        \caption{The relationship between the number of customers in the transaction data set, the number of users in the Twitter data set, and the neighborhood-level wealth, for the European metropolitan area.}
        \label{fig:figS1}
\end{figure}

\begin{figure}
      \centering
      {\includegraphics[width=16cm]{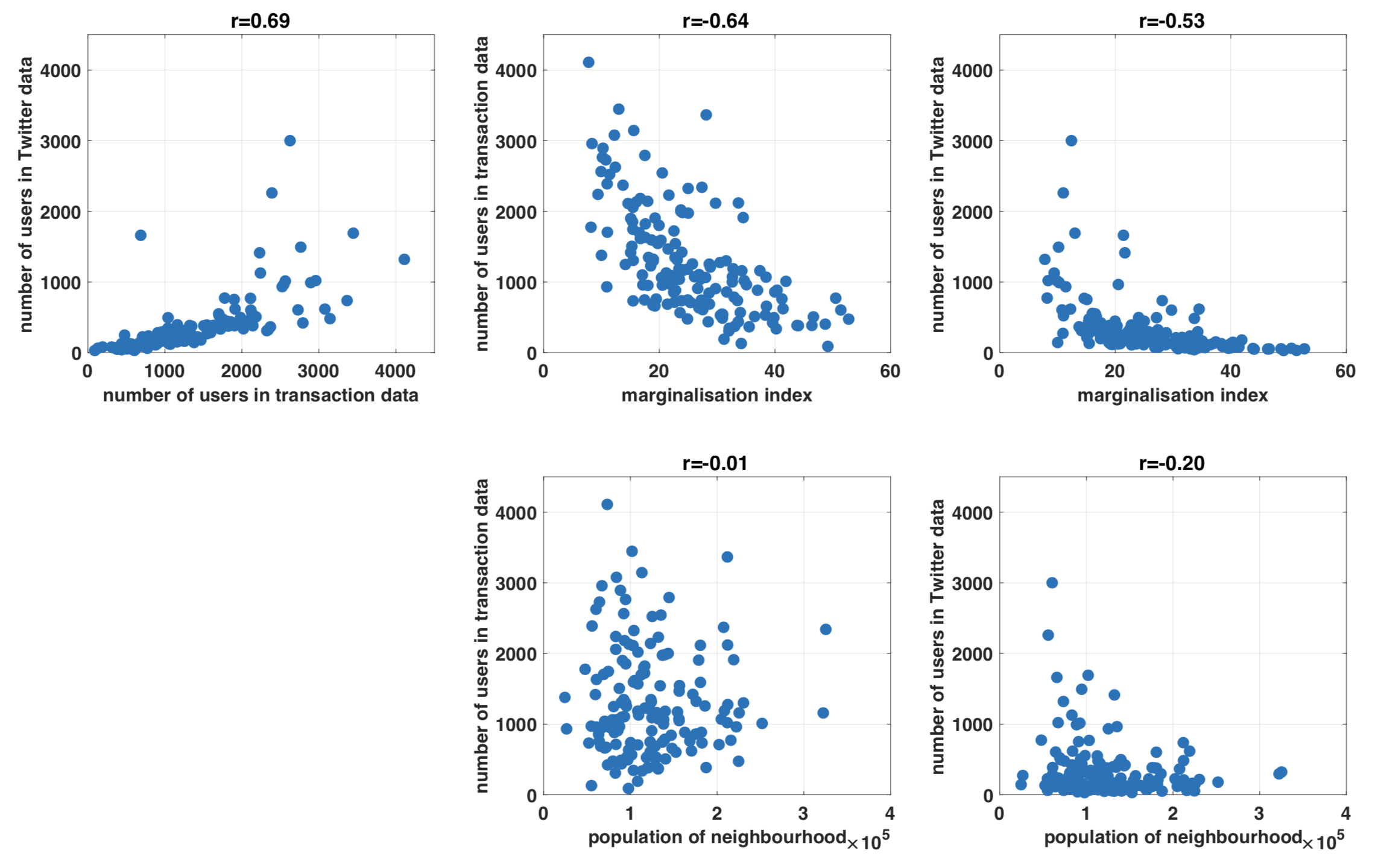}}
        \caption{The relationship between the number of customers in the transaction data set, the number of users in the Twitter data set, and the neighborhood-level wealth, for the Latin American metropolitan area.}
        \label{fig:figS2}
\end{figure}

\begin{figure}
      \centering
      {\includegraphics[width=16cm]{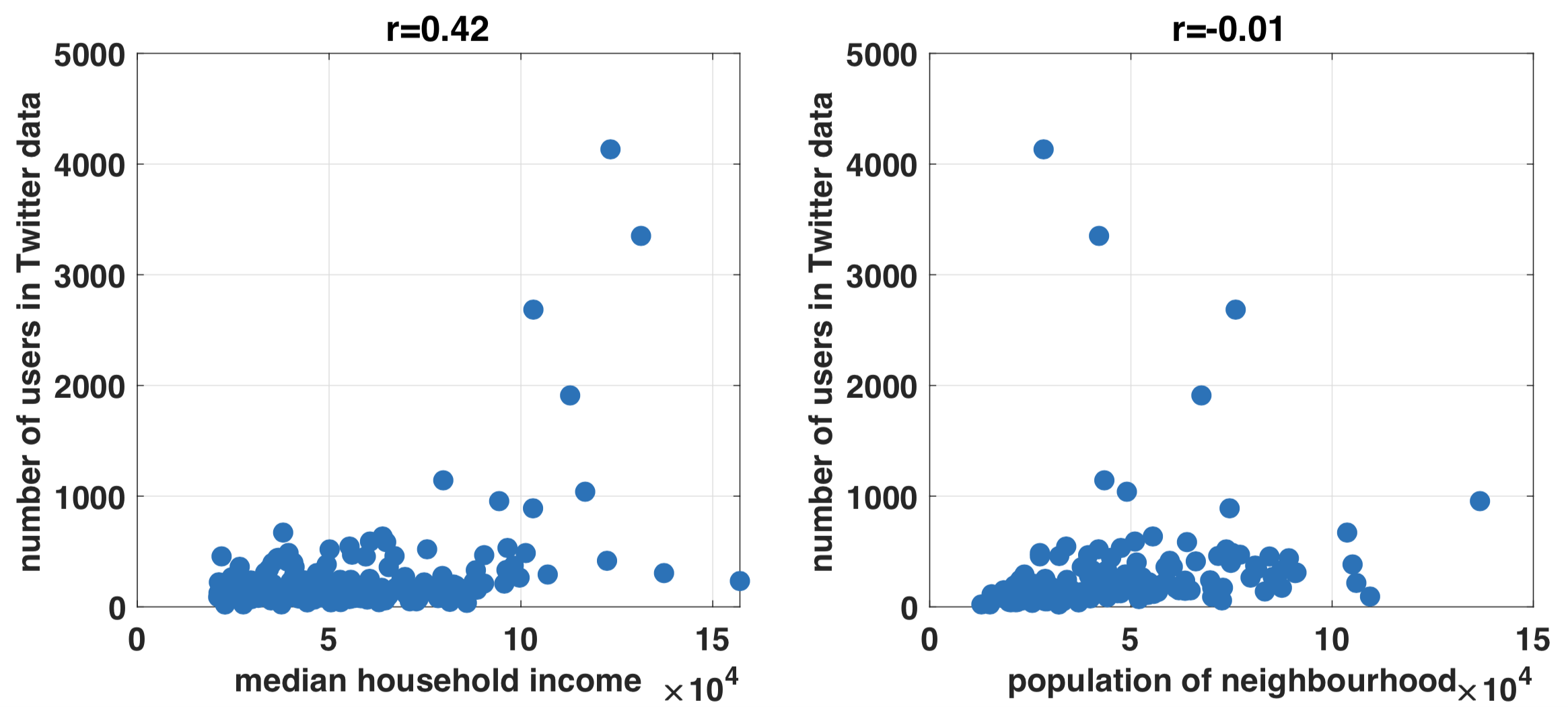}}
        \caption{The relationship between the number of users in the Twitter data set and the neighborhood-level wealth, for the Northern American metropolitan area.}
        \label{fig:figS3}
\end{figure}

{\subsection*{Data statistics}
Figure~\ref{fig:figS1} (Top Left) and Figure~\ref{fig:figS2} (Top Left) illustrate the comparison between the number of credit card customers and Twitter users in the European and in the Latin American cases, respectively.
We see that, for both metropolitan areas, the number of people in the two data sets are correlated ($r=0.76$ {in the European case and $r=0.69$ in the Latin America case}).
The other plots in Figure~\ref{fig:figS1}, Figure~\ref{fig:figS2} and Figure~\ref{fig:figS3} show the relationship between the number of credit card users, number of Twitter users, the neighborhood population, and the neighborhood-level socio-economic status, for the three metropolitan areas, respectively.
It can be seen that in the European case the number of users in the credit card and Twitter data is strongly correlated with population ($r=0.74$ and $r=0.62$), and not biased towards population with certain socio-economic status ($r=0.25$ and $r=0.24$). In comparison, the correlation between the number of credit card and Twitter users and the population is much weaker for the Latin American ($r=-0.01$ and $r=-0.20$) and Northern American ($r=-0.01$) cases, where the sampling of users is biased towards richer neighborhoods ($r=-0.64$ and $r=-0.53$ for the Latin American case and $r=0.42$ for the Northern American case).
To address the issue of bias in sampling, in particular for the Latin American and Northern American cases, we construct population-weighted interaction networks as described below.}

\begin{figure}
      \centering
      {\includegraphics[width=16cm]{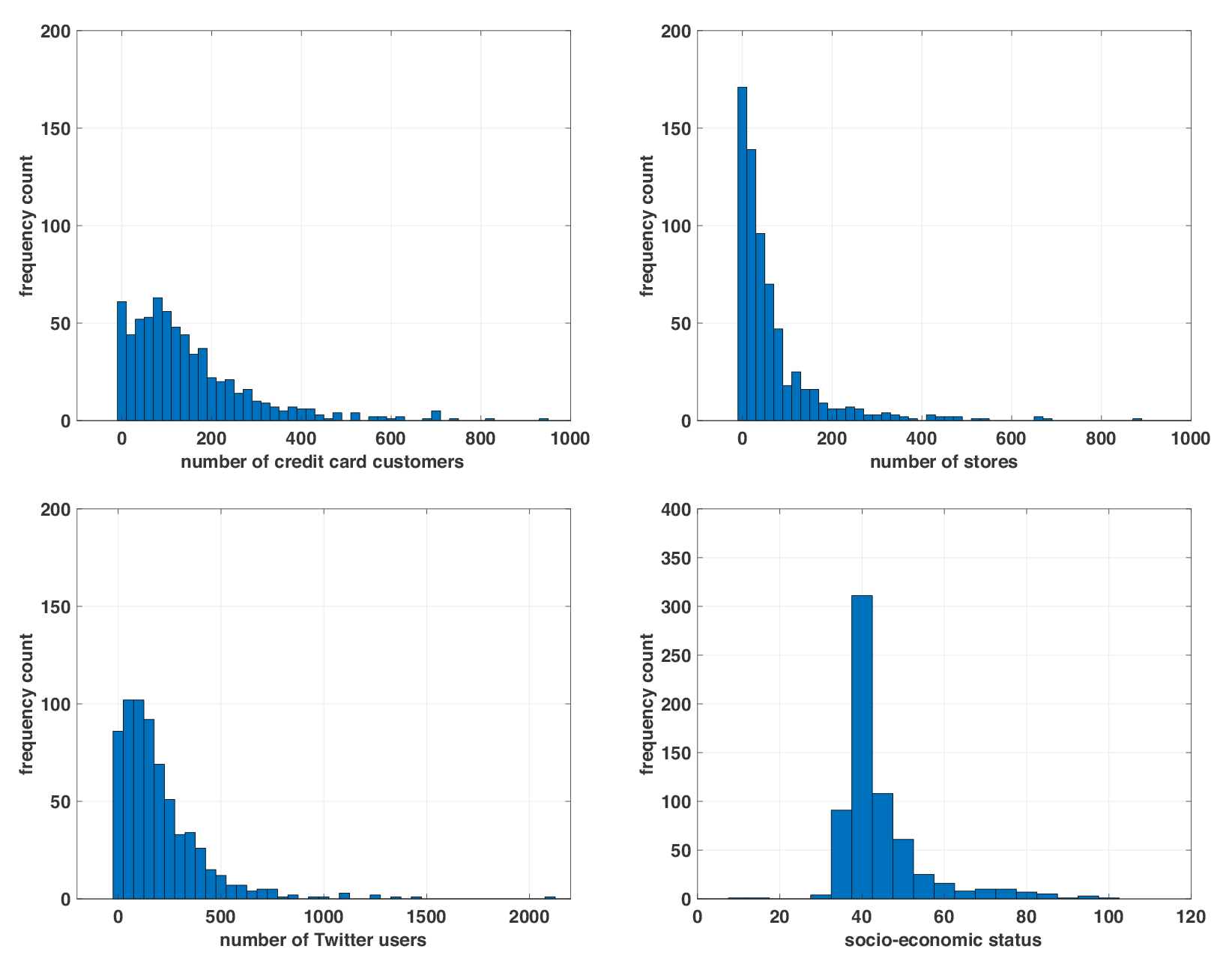}}
        \caption{Histograms of (Top Left) the number of credit card customers, (Top Right) the number of stores, (Bottom Left) the number of Twitter users, and (Bottom Right) the socio-economic status, for neighborhoods in the European metropolitan area.}
        \label{fig:figS4}
\end{figure}

\begin{figure}
      \centering
      {\includegraphics[width=16cm]{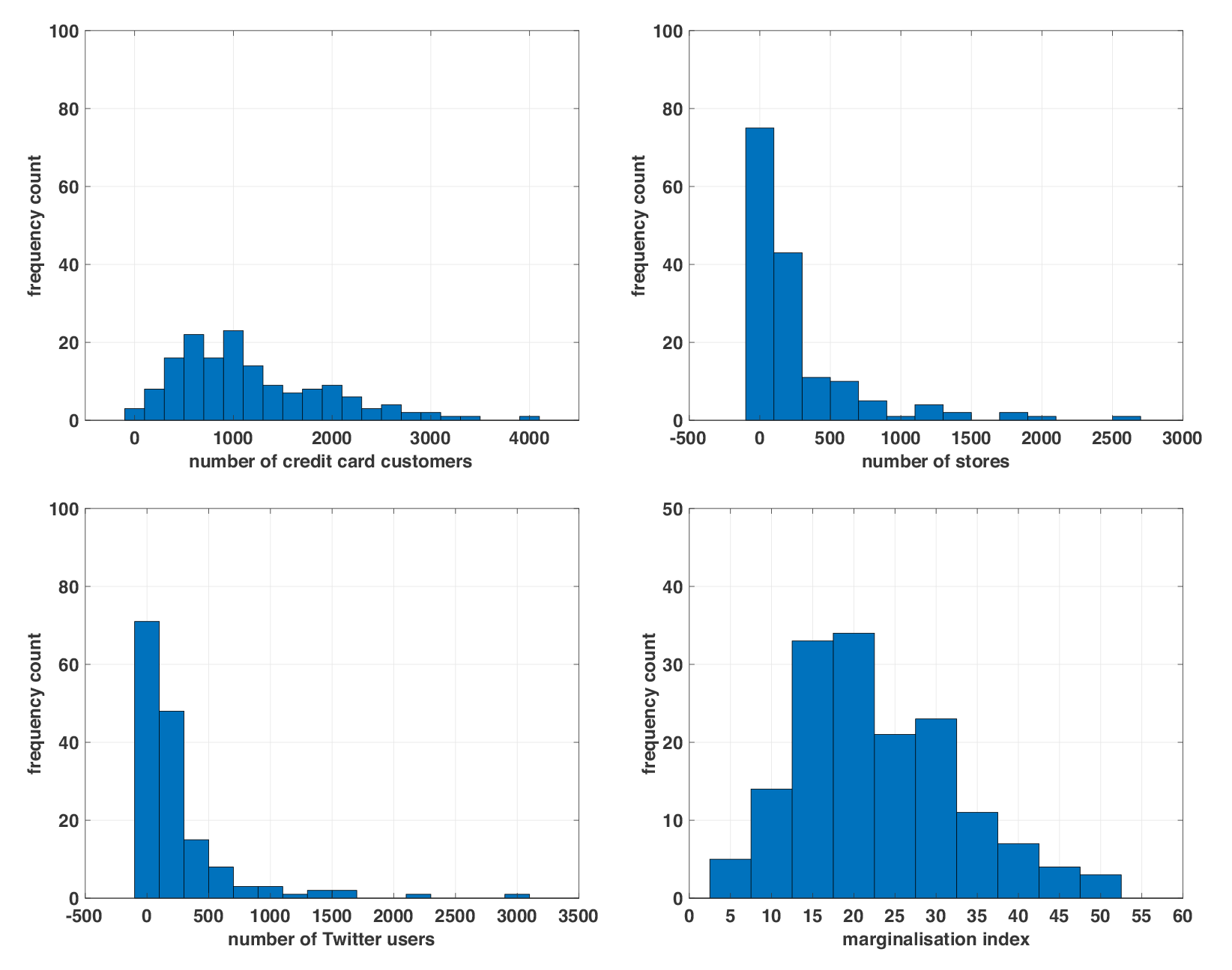}}
        \caption{Histograms of (Top Left) the number of credit card customers, (Top Right) the number of stores, (Bottom Left) the number of Twitter users, and (Bottom Right) the socio-economic status, for neighborhoods in the Latin American metropolitan area.}
        \label{fig:figS5}
\end{figure}

\begin{figure}
      \centering
      {\includegraphics[width=16cm]{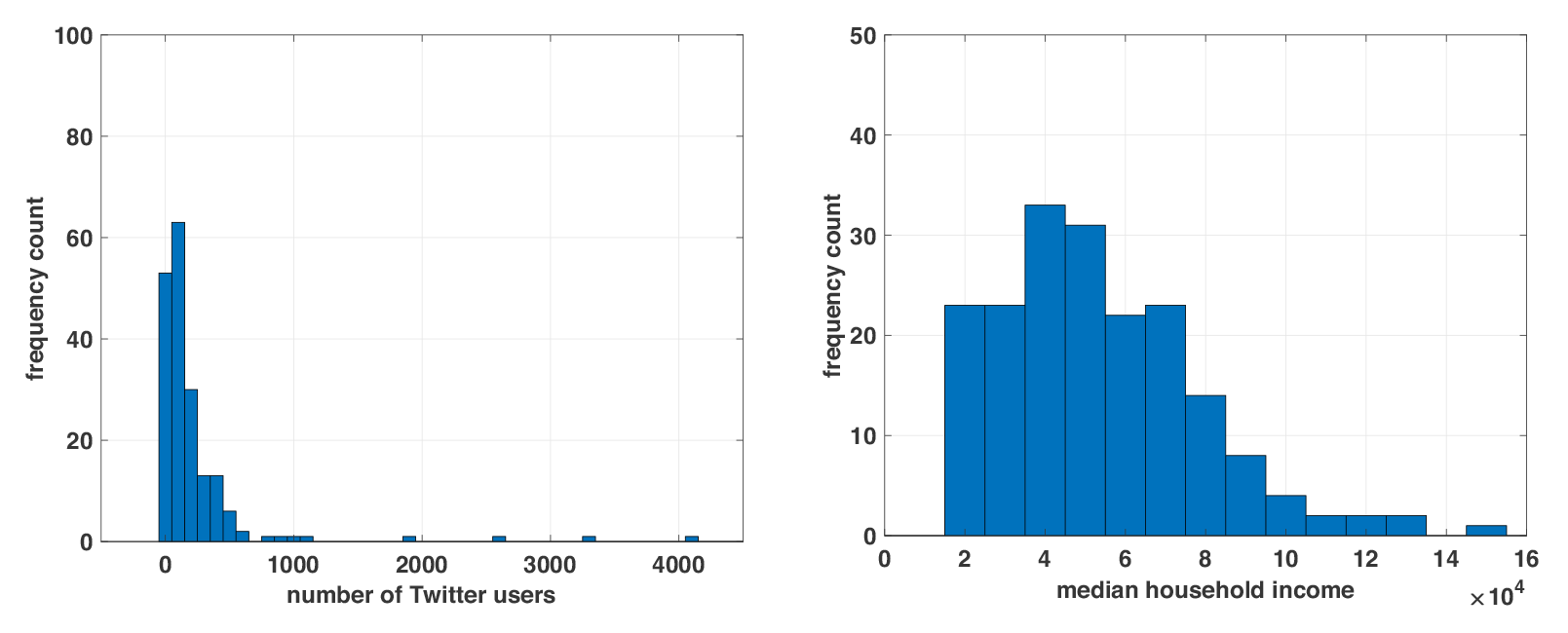}}
        \caption{Histograms of (Left) the number of Twitter users, and (Right) the median household income, for neighborhoods in the Northern American metropolitan area.}
        \label{fig:figS6}
\end{figure}

{Figure~\ref{fig:figS4} shows the distribution of the number of credit card customers, stores, Twitter users, and that of the socio-economic status of the neighborhoods in the European metropolitan area, where Figure~\ref{fig:figS5} and Figure~\ref{fig:figS6} show the same distribution in the Latin American and Northern American metropolitan areas, respectively.}

\subsection*{Computation of the diversity score}
For each individual $s$, we define the diversity as the Shannon entropy of his/her purchase (or Twitter) activities:
\begin{equation}
D(s) = -\sum_{t=1}^N p_{st}~\text{log}(p_{st}),
\end{equation}
where $p_{st}$ is the probability that an individual $s$ (or Twitter user) visits a store $t$ (or mentions another user $t$) and $N$ is the total number of stores (or Twitter users).
The average diversity score of a neighborhood is then defined as the average diversity of individuals living in that neighborhood. This approach is similar to the network-based approach of Eagle \emph{et al.} \cite{Eagle10b}.

\subsection*{Construction of the interaction networks}
We construct two networks to capture interactions between different neighborhoods. In these networks, nodes represent neighborhoods and edges represent interactions whose intensities are captured by the weights of the edges.
For purchase network, we define a directional edge from neighborhood $i$ to $j$ with weight $w_{ij}^{(p)}$, which is the number of purchases made by customers living in $i$ at stores in $j$.
Similarly, for Twitter mention network, we define a directional edge from neighborhood $i$ to $j$ with weight $w_{ij}^{(t)}$, which is the number of mentions made by Twitter users living in $i$ to users in $j$.

{\subsection*{Construction of the population-weighted interaction networks}
To address the issue of bias in the sampling of the users in the credit card and Twitter data sets, we scale the interaction networks using a population based weighting scheme. More specifically, for the credit card data set, if a neighbourhood $i$ has $m_i$ credit card users and $p_i$ population, and the number of purchases from $i$ to $j$ is $w_{ij}^{(p)}$, then we define the population-weighted interaction in the credit card case as:
\begin{equation}
\bar{w}_{ij}^{(p)} = \frac{w_{ij}^{(p)}}{m_i / p_i}.
\end{equation}
Similarly, for the Twitter data set, if neighbourhood $i$ has $x_i$ twitter users and $p_i$ population, neighbourhood $j$ has $m_j$ twitter users and $p_j$ population, and the number of mentions from $i$ to $j$ is $w_{ij}^{(t)}$, then we define the population-weighted interaction in the Twitter case as:
\begin{equation}
\bar{w}_{ij}^{(t)} = \frac{w_{ij}^{(t)}}{(m_i \times m_j) / (p_i \times p_j)}.
\end{equation}}

\subsection*{Construction of the mixing matrices}
\begin{figure}
      \centering
      {\includegraphics[width=16cm]{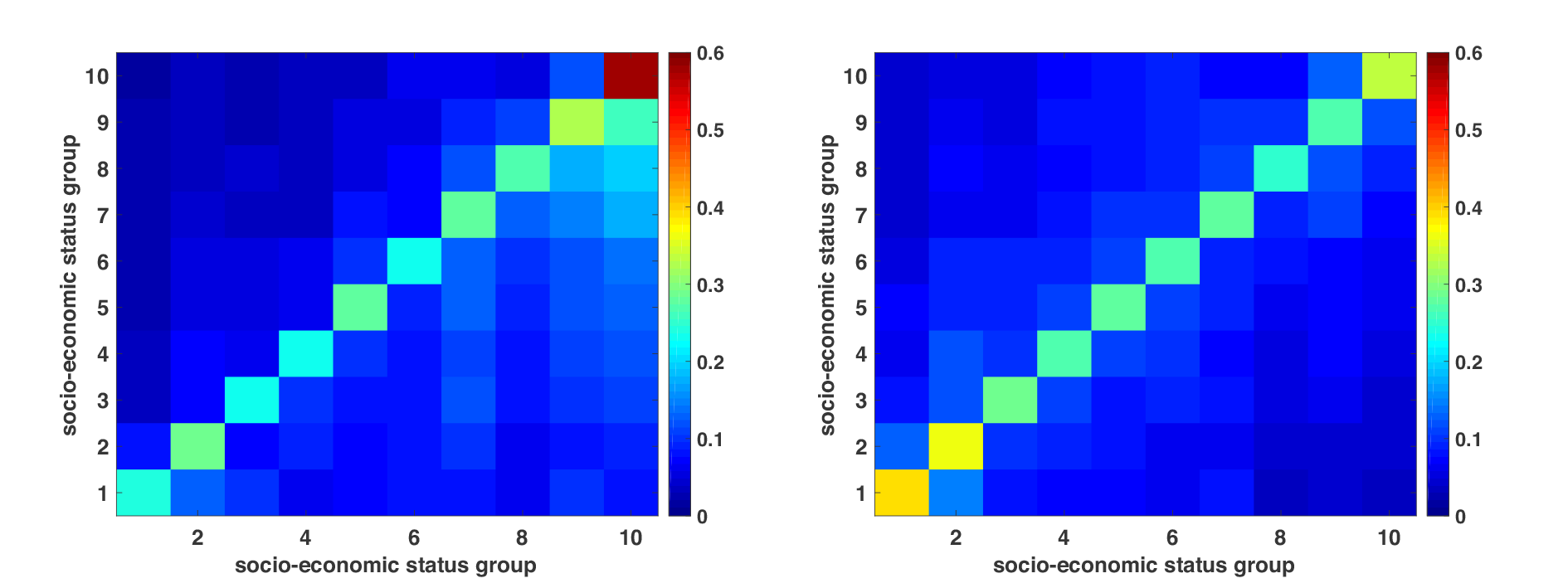}}
        \caption{Mixing matrices for (Left) the purchase network and (Right) the {Twitter} mention network, for ten socio-economic status groups in the European metropolitan area. Socio-economic status groups are ordered from the lowest wealth (1) to the highest wealth (10).}
        \label{fig:figS7}
\end{figure}

\begin{figure}
      \centering
      {\includegraphics[width=16cm]{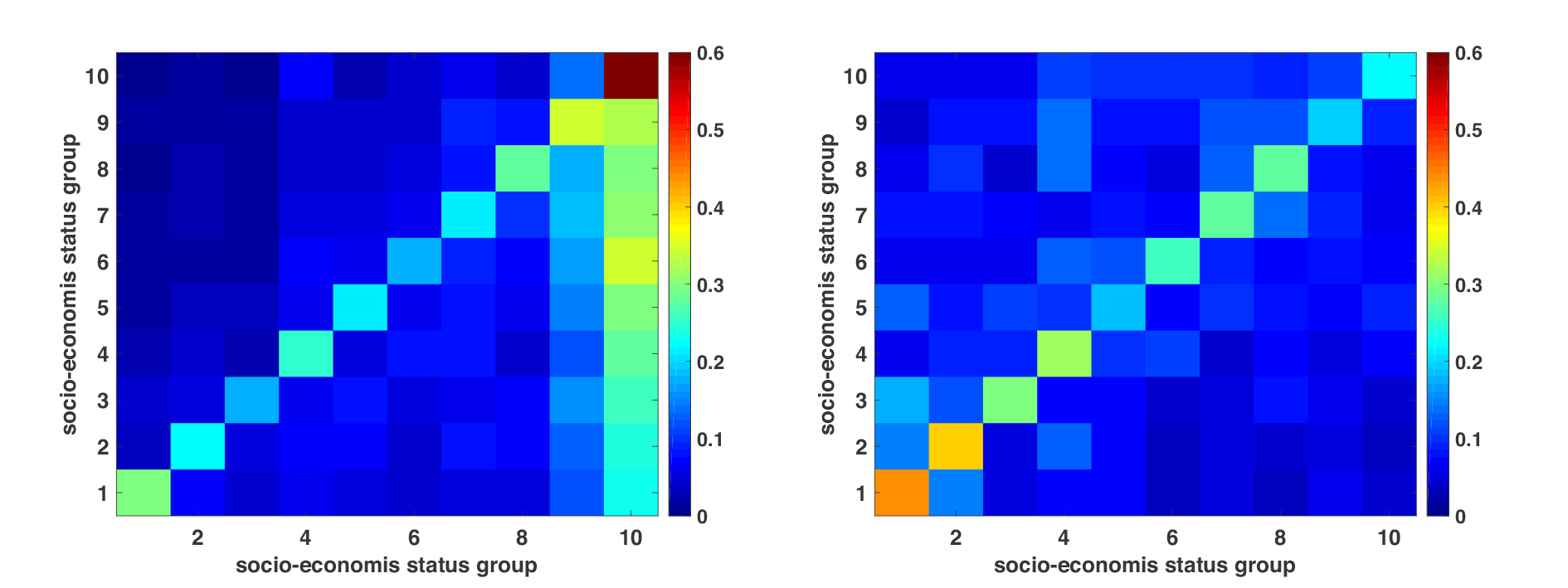}}
        \caption{Mixing matrices for the (Left) purchase network and (Right) mention network, for ten socio-economic status groups, in the Latin American metropolitan area. Socio-economic status groups are ordered from the lowest wealth (1) to the highest wealth (10).}
        \label{fig:figS8}
\end{figure}

\begin{figure}
      \centering
      {\includegraphics[width=8cm]{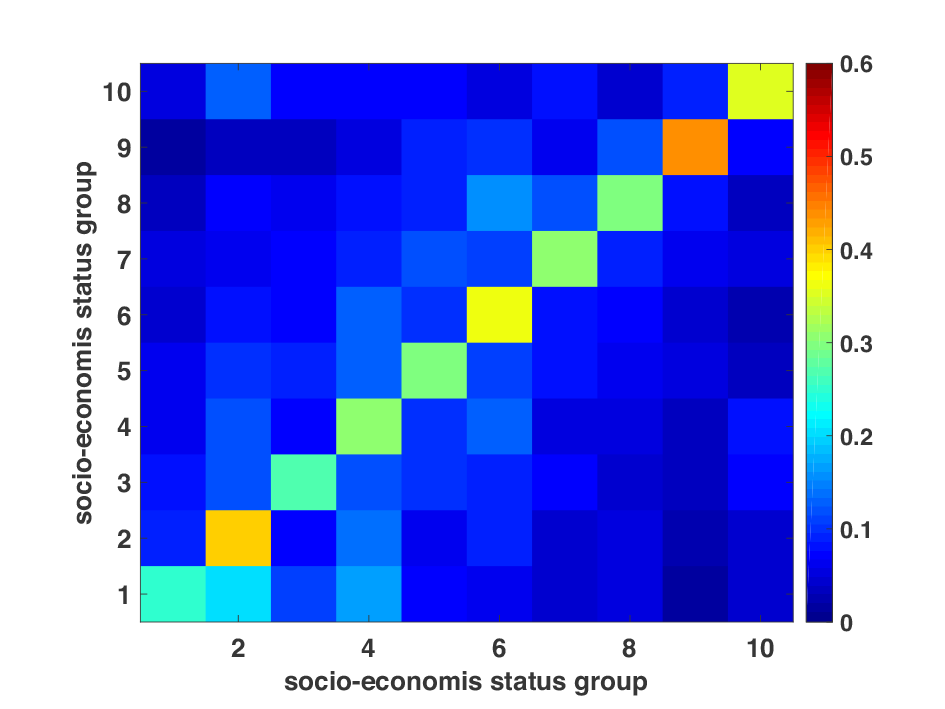}}
        \caption{Mixing matrix for the mention network, for ten income groups, in the Northern American metropolitan area. Socio-economic status groups are ordered from the lowest wealth (1) to the highest wealth (10).}
        \label{fig:figS9}
\end{figure}

To quantify segregation, we first construct the mixing matrices. {Specifically, we put the neighborhoods into ten groups according to their socioeconomic status, where all groups have an equal number of neighborhoods. The groups have increasing socio-economic status from 1 to 10, i.e., the socio-economic status group for neighborhood $i$ is $s(i) = [1,2,\dots,10]$.} We then convert the population-weighted interaction networks into the 10 by 10 mixing matrices, whose $mn$-th entry is defined as:
\begin{equation}
\begin{split}
M_{mn}^{(p)} &= \sum_{s(i)=m, s(j)=n} \bar{w}_{ij}^{(p)},\\
M_{mn}^{(t)} &= \sum_{s(i)=m, s(j)=n} \bar{w}_{ij}^{(t)}.
\end{split}
\end{equation}
Finally, we normalize the mixing matrix of both networks into a stochastic matrix:
\begin{equation}
\begin{split}
S_{mn}^{(p)} = \frac{ M_{mn}^{(p)} }{ \sum_n M_{mn}^{(p)} },\\
S_{mn}^{(t)} = \frac{ M_{mn}^{(t)} }{ \sum_n M_{mn}^{(t)} },
\end{split}
\end{equation}
This way, $S_{mn}^{(p)}$ and $S_{mn}^{(t)}$ represent the probability of interaction from one socio-economic status group $m$ to another $n$ in terms of credit card purchases and Twitter mentions.
The resulting mixing matrices for the interaction networks in the European, {Latin American, and Northern American case} are shown in Figure~\ref{fig:figS7}, {Figure~\ref{fig:figS8}, and Figure~\ref{fig:figS9}, respectively}.

\subsection*{Computation of the assortative mixing coefficient}
Assortative mixing coefficient (or assortativity) is a measure proposed by Newman \emph{et al.} \cite{Newman03} to quantify the phenomenon of homophily in social networks, which can also be used for measuring segregation in networks \cite{Rodriguez16}. {In the context of the mixing matrices, it is equivalent to Cohen's Kappa, a classical psychometric measure of agreement on nominal variables \cite{Bojanowskia14}.}
In our case, the more assortative the {mixing} matrices, the more segregated the behavioral interaction patterns.

The assortativity proposed in \cite{Newman03} is computed as follows. Given a weighted network, let $e_{xy}$ be the fraction of weights of edges in the network that join nodes having attribute values $x$ and $y$. The assortativity is then defined as:
\begin{equation}
r = \frac{\sum_{xy} xy (e_{xy}-a_x b_y)}{\sigma_x \sigma_y},
\label{eq:amc}
\end{equation}
where $\sum_{xy}e_{xy}=1$,  $a_x=\sum_{y}e_{xy}$ is the fraction of weights of edges starting from nodes with attribute value $x$, $b_y=\sum_{x}e_{xy}$ is the fraction of that connecting to nodes with attribute value $y$, and 
\begin{equation}
\sigma_x = \sqrt{\sum_x x^2 a_x - \big(\sum_x x a_x\big)^2}
\end{equation}
and
\begin{equation}
\sigma_y = \sqrt{\sum_y y^2 b_y - \big(\sum_y y b_y\big)^2}
\label{eq:amc}
\end{equation}
are the standard deviations of distributions of $a_x$ and $b_y$, respectively. As we can see, the assortativity $r$ is the Pearson correlation coefficient between the attributes of the two end nodes for all the edges.

{In Appendix, we describe in detail how we compute the assortativity based on a percentage of socio-economic status groups as well as geographical distance between neighborhoods.}

\section*{Acknowledgements}
The authors would like to thank the European bank as well as the American company who donated the credit card transaction data sets for this research.

\bibliography{EPJ}
\bibliographystyle{Science}

\clearpage
\section*{Appendix}

\subsection*{Computation of assortativity based on a percentage of socio-economic status groups}
\begin{figure}
      \centering
      {\includegraphics[width=16cm]{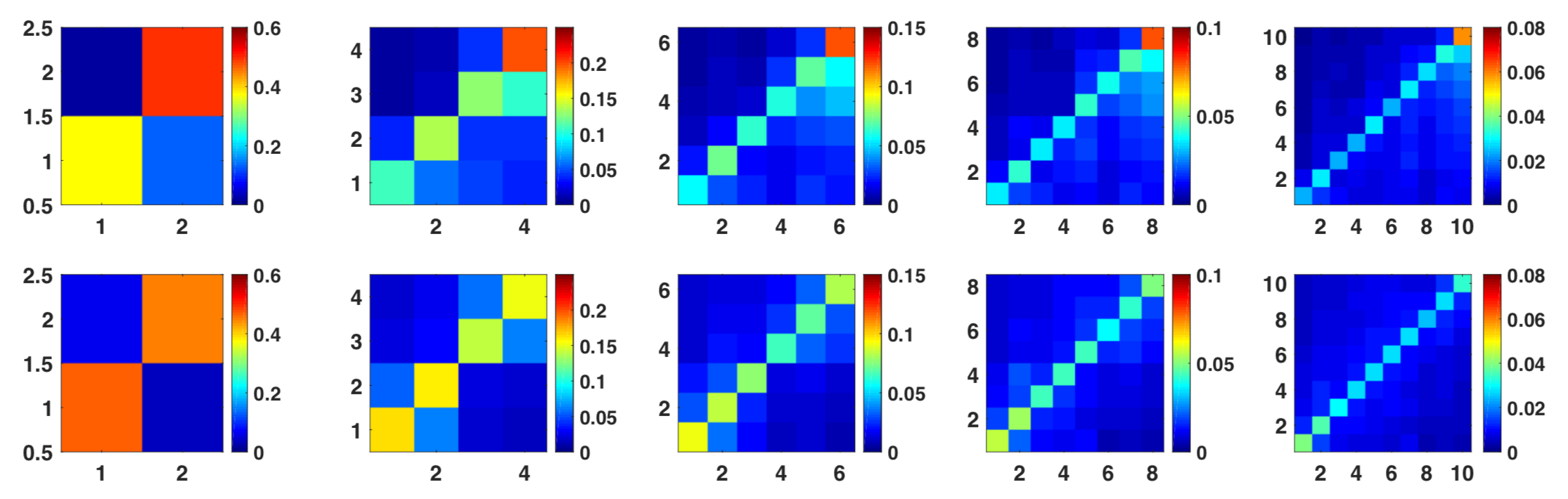}}
        \caption{Mixing matrices of (Top) purchase and (Bottom) Twitter mention networks for extreme neighborhoods in terms of socio-economic status group for the European metropolitan area. {The columns from left to right correspond to the consideration of 1, 2, 3, 4 and 5 extreme groups on both the low and high socio-economic distribution.}}
        \label{fig:figS10}
\end{figure}

\begin{figure}
      \centering
      {\includegraphics[width=16cm]{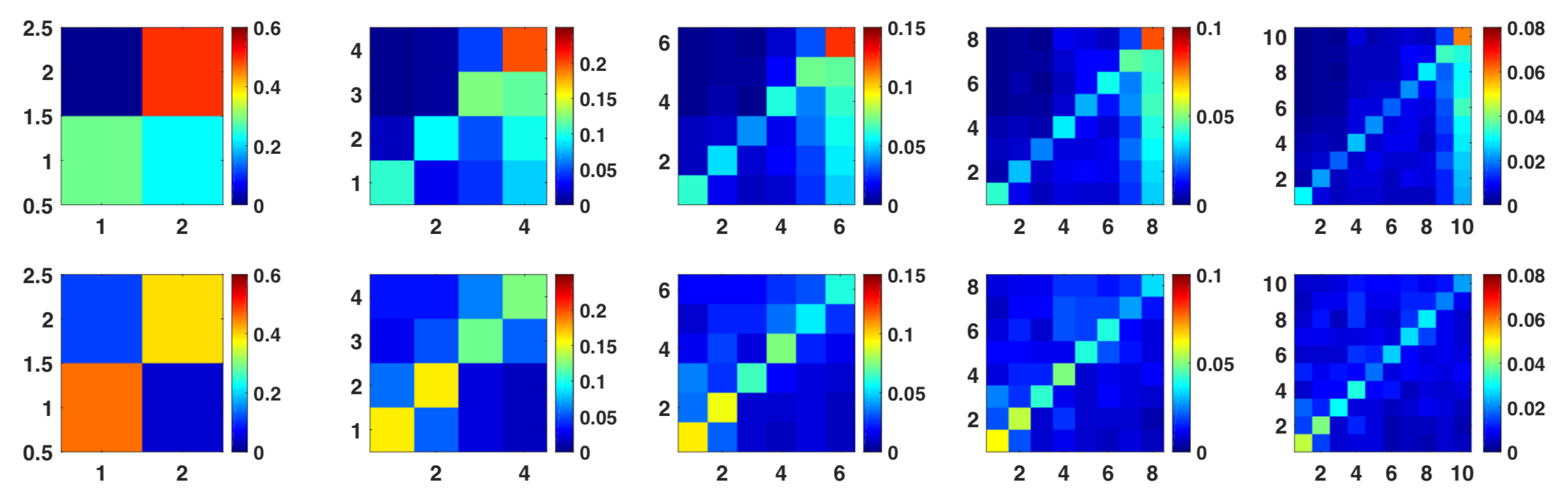}}
        \caption{Mixing matrices of (Top) purchase and (Bottom) Twitter mention networks for extreme neighborhoods in terms of socio-economic status group for the Latin American metropolitan area. {The columns from left to right correspond to the consideration of 1, 2, 3, 4 and 5 extreme groups on both the low and high socio-economic distribution.}}
        \label{fig:figS11}
\end{figure}

\begin{figure}
      \centering
      {\includegraphics[width=16cm]{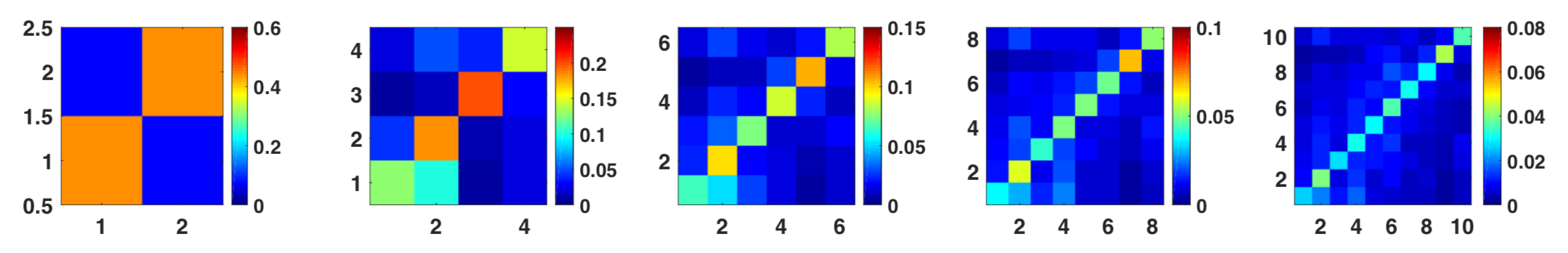}}
        \caption{Mixing matrices of Twitter mention networks for extreme neighborhoods in terms of median household income for the Northern American metropolitan area. {The columns from left to right correspond to the consideration of 1, 2, 3, 4 and 5 extreme groups on both the low and high socio-economic distribution.}}
        \label{fig:figS12}
\end{figure}

We propose the following framework for analyzing segregation among groups of neighborhoods of different socio-economic status. We first consider only entries in $M_{mn}^{(p)}$ and $M_{mn}^{(t)}$ that correspond to groups 1 and 10 (those of lowest and highest wealth), and normalize the resulting 2 by 2 matrices as described above to obtain the mixing matrices that correspond to only these two socio-economic status groups (shown in the first column of Figure~\ref{fig:figS10}, Figure~\ref{fig:figS11} and Figure~\ref{fig:figS12}). We then compute the assortativity of these mixing matrices. 
Next, we include groups 2 and 9, and compute the 4 by 4 mixing matrices as well as the assortativity. This process is repeated until we eventually include all socio-economic status groups. The mixing matrices at each step are shown in the different columns of Figure~\ref{fig:figS10}, Figure~\ref{fig:figS11} and Figure~\ref{fig:figS12} for the European, Latin American, and Northern American cases, respectively. These are the matrices based on which we produce the results shown in Figure~\ref{fig:fig3}.

\subsection*{Computation of assortativity based on geographical distance}
For analyzing interaction patterns between neighborhood pairs of different geographical distances, we propose the following framework.
We first prune the interaction networks by removing edges that correspond to neighborhood pairs of distance smaller or larger than a set of thresholds. We then compute the assortativity that corresponds to the networks with the remaining subset of edges.

The distance thresholds are chosen to be the 20, 40, 60, 80, and 100 percentiles of a vector containing all the pairwise distances, which are 8.5km, 15km, 22km, 31km and 85km in the European case, 12km, 18km, 25km, 34km and 91km in the Latin American case, and 8.1km, 13km, 18km, 23km and 52km in the Northern American case. These thresholds are generally consistent across the three metropolitan areas.

The results are shown in the bottom row of Figure~\ref{fig:fig3}. It is natural to see that keeping edges greater then 0km (yellow and purple curves in Figure~\ref{fig:fig3} (bottom left)) is equivalent to removing edges greater than 85km (blue and orange curves in Figure~\ref{fig:fig3} (bottom left)), which is the maximum distance between any pair of neighborhoods.

\subsection*{Construction of simulated interaction networks using a gravity-based model}
To illustrate the difference between the empirical interaction patterns and the one that would have been caused by geographic distribution of neighborhoods,
{we simulate offline and online interaction networks between neighborhoods by considering the following model similar to the gravity-based model considered in \cite{Krings09}:
\begin{equation}
\begin{split}
w_{ij}^{(p)} &\approx c_p \frac{ {[n_{i}^{(p)}]}^{\beta_{p1}} ~ {[m_{j}^{(p)}]}^{\beta_{p2}} }{ {[T_{ij}+\epsilon_p]}^{\alpha_p} },\\
w_{ij}^{(t)} &\approx c_t \frac{ {[n_{i}^{(t)}]}^{\beta_{t1}} ~ {[m_{j}^{(t)}]}^{\beta_{t2}} }{ {[T_{ij}+\epsilon_t]}^{\alpha_t} },
\end{split}
\label{eq:fitting}
\end{equation}
where $w_{ij}^{(p)}$ and $w_{ij}^{(t)}$ are the empirically observed edge weights in the purchase and Twitter networks, respectively, $n_{i}^{(p)}$ and $n_{i}^{(t)}$ are the numbers of credit card customers and Twitter users in neighborhood $i$, respectively, and $T_{ij}$ is the geographical distance between the centroids of the neighborhoods $i$ and $j$.}

\begin{figure}
      \centering
      {\includegraphics[width=16cm]{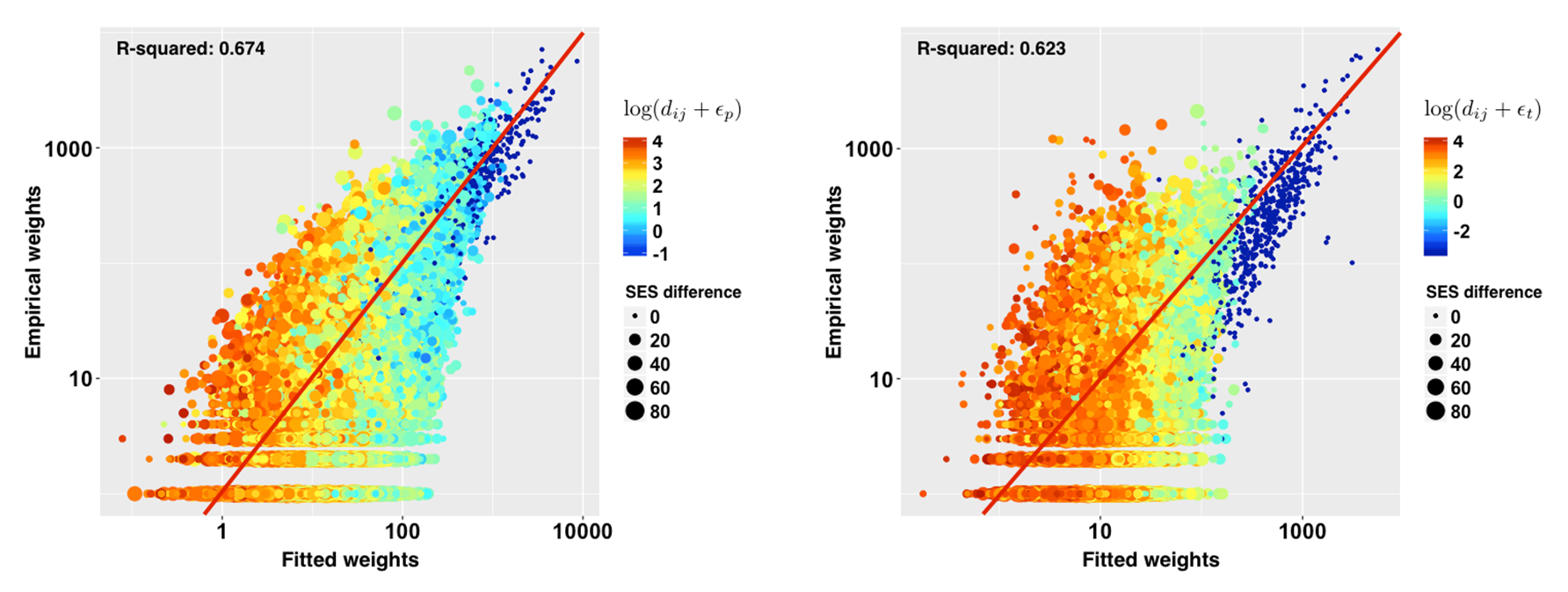}}
        \caption{Weighted OLS fitting for (Left) number of purchases and (Right) number of mentions, for the European metropolitan area.}
        \label{fig:figS13}
\end{figure}

\begin{figure}
      \centering
      {\includegraphics[width=16cm]{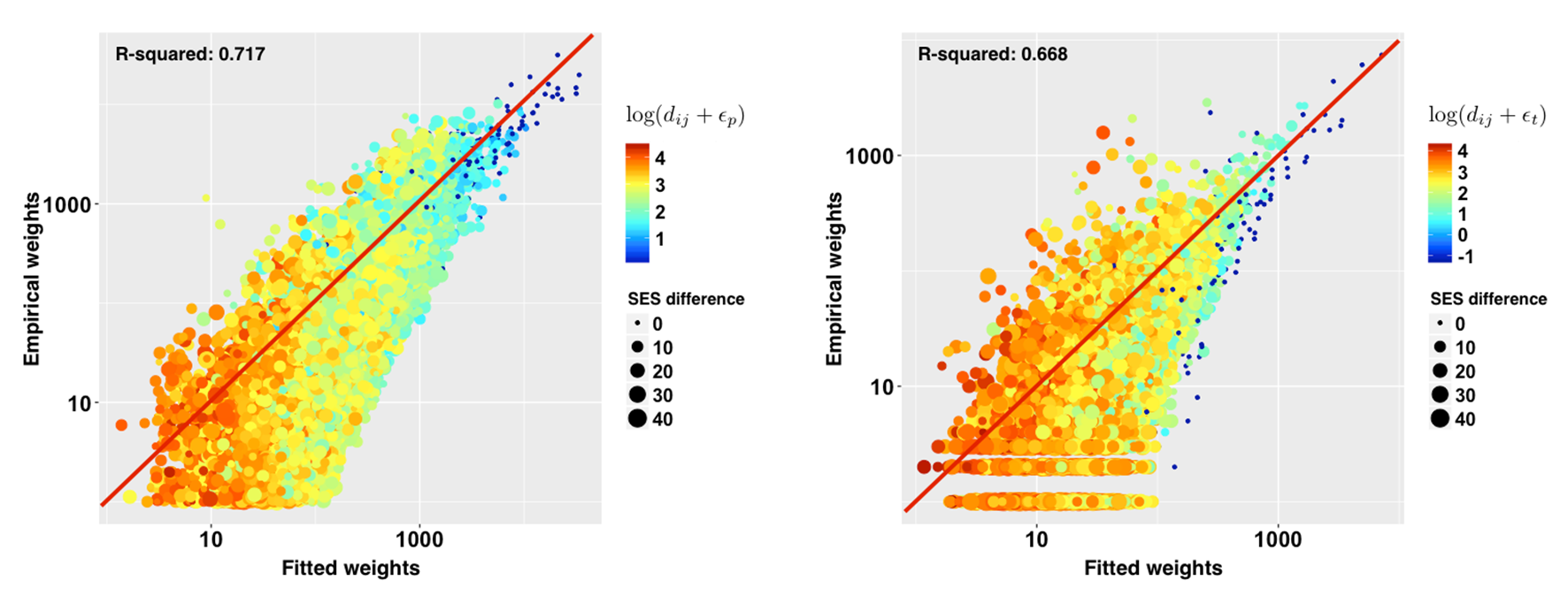}}
        \caption{Weighted OLS fitting for (Left) number of purchases and (Right) number of mentions, for the Latin American metropolitan area.}
        \label{fig:figS14}
\end{figure}

\begin{figure}
      \centering
      {\includegraphics[width=8cm]{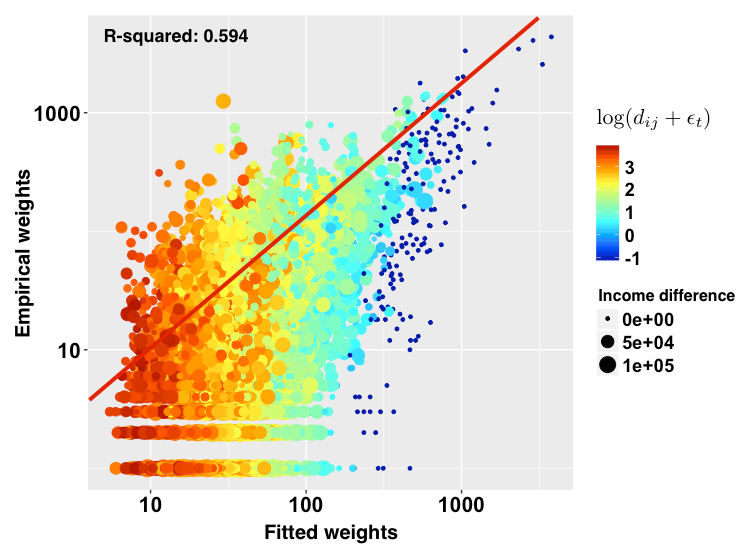}}
        \caption{Weighted OLS fitting for number of mentions, for the Northern American metropolitan area.}
        \label{fig:figS15}
\end{figure}

We obtain optimal values for the parameters $c_p$, $\beta_{p1}$, $\beta_{p2}$, $\epsilon_p$, $\alpha_p$, $c_t$, $\beta_{t1}$, $\beta_{t2}$, $\epsilon_t$ and $\alpha_t$ by fitting a weighted Ordinary Least Squares (OLS) model to the observed number of purchases $w_{ij}^{(p)}$ or mentions $w_{ij}^{(t)}$, where we weight the fitting for $w_{ij}^{(p)}$ or $w_{ij}^{(t)}$ with its own value:
\begin{equation}
\begin{split}
\text{log}~w_{ij}^{(p)} &\approx \big( \text{log}~c_p + \beta_{p1}~n_{i}^{(p)} + \beta_{p2}~m_{j}^{(p)} - \alpha_p~(T_{ij}+\epsilon_p) \big) \times w_{ij}^{(p)},\\
\text{log}~w_{ij}^{(t)} &\approx \big( \text{log}~c_t + \beta_{t1}~n_{i}^{(t)} + \beta_{t2}~m_{j}^{(t)} - \alpha_t~(T_{ij}+\epsilon_t) \big) \times w_{ij}^{(t)}.
\end{split}
\label{eq:fitting2}
\end{equation}
The values of the parameters for the three cities are shown in Table~\ref{tab:params}, and the fitting for the two interaction networks in each city is shown in Figure~\ref{fig:figS13}, Figure~\ref{fig:figS14} and Figure~\ref{fig:figS15}, respectively.

Upon obtaining these parameters, we compute simulated number of purchases and mentions $\hat{w}_{ij}^{(p)}$ and $\hat{w}_{ij}^{(t)}$ using the right hand side of Equation~(\ref{eq:fitting}).
{We then scale these simulated networks using the population based weighting scheme described in Materials and Methods.} 
Finally, we compute the mixing matrices and the segregation index (assortativity) according to the gravity-based model.
The results are shown in the top row of Figure~\ref{fig:fig3} for the European, Latin American, and Northern American cases.

\subsection*{Construction of artificial interaction networks using a null model}
We further validate the observed segregation pattern by comparing it against {the one produced} by a null model, in which socio-economic status of the neighborhoods are randomized to leave only the segregation effect of individuals visiting stores or mentioning others in their home neighborhoods.
Specifically, we randomly shuffle the socio-economic status of the neighborhoods, and construct the artificial interaction networks for both offline purchases and online {Twitter} mentions. {Notice that these artificial interaction networks have the same pairwise edge weights as in the empirical interaction networks, but with node attribute (socio-economic status of the neighborhoods) randomly shuffled. We then scale these artificial interaction networks using the population based weighting scheme described in Materials and Methods.}
Finally, we compute the mixing matrices and the segregation index (assortativity) according to the null model.
The results are shown in the top row of Figure~\ref{fig:fig3} for the European, Latin American, and Northern American cases.

\subsection*{The jackknife resampling}
To test the sensitivity of the assortativity of the interaction networks to certain edges, we use the jackknife resampling technique originally proposed in \cite{Efron79} and then adopted in \cite{Newman03}. The idea is to randomly remove a certain percentage (5\% in our case) of edges in each network, and then re-compute the assortativity of the network. For all the results shown in Figure~\ref{fig:fig3}, we apply the jackknife resampling for 100 times, and compute the 95\% confidence interval of the assortativity.

\subsection*{Segregation and economic inequality}
Segregation by income has direct implications on emerging economic inequalities, such as the unequal flow of money in the city \cite{Greenwood14}. We investigate this relationship by analyzing the flow of money in the city using the purchase behavior of individuals and the distribution of sales revenue across neighborhoods. To this end, we add up the sales revenue of all the stores in each neighborhood, and compute the GINI coefficient of the resulting distribution across neighborhoods, similarly to the previous study in \cite{Louail17}.
The GINI coefficient can be thought of as an approximation of the economic inequality between the neighborhoods, which is then analyzed together with the assortativity (segregation index). 

In addition to the empirical purchase networks, we repeat the same analysis on two artificial networks: (i) simulated purchase networks based on a gravity model {(as in the other analyses)} and (ii) networks produced by randomly reshuffling the location (in terms of neighborhood) of a fraction of the stores as well as customer homes in a null model.
For the simulated network based on the gravity model, we adjust each transaction amount from neighborhood $i$ to $j$ by multiplying the transaction amount with the ratio of the actual number of transactions from $i$ to $j$ to the simulated one. {We then scale all the networks using the population based weighting scheme described in Materials and Methods.}
Finally, we compute the GINI coefficient as well as the assortativity corresponding to all the networks.

For the networks based on random reshuffling, we use five fractions (i.e., $20\%$, $40\%$, $60\%$, $80\%$ and $100\%$) for the reshuffling procedure.
It is worth noting that such reshuffling does not change the number of customers and stores in each neighborhood nor the amount for each transaction, but only breaks the segregation pattern in purchase behavior. For each fraction, we apply the random reshuffling procedure for 50 times to compute the standard deviation of the resulting GINI coefficient and assortativity.

\begin{figure}
      \centering
      {\includegraphics[width=16cm]{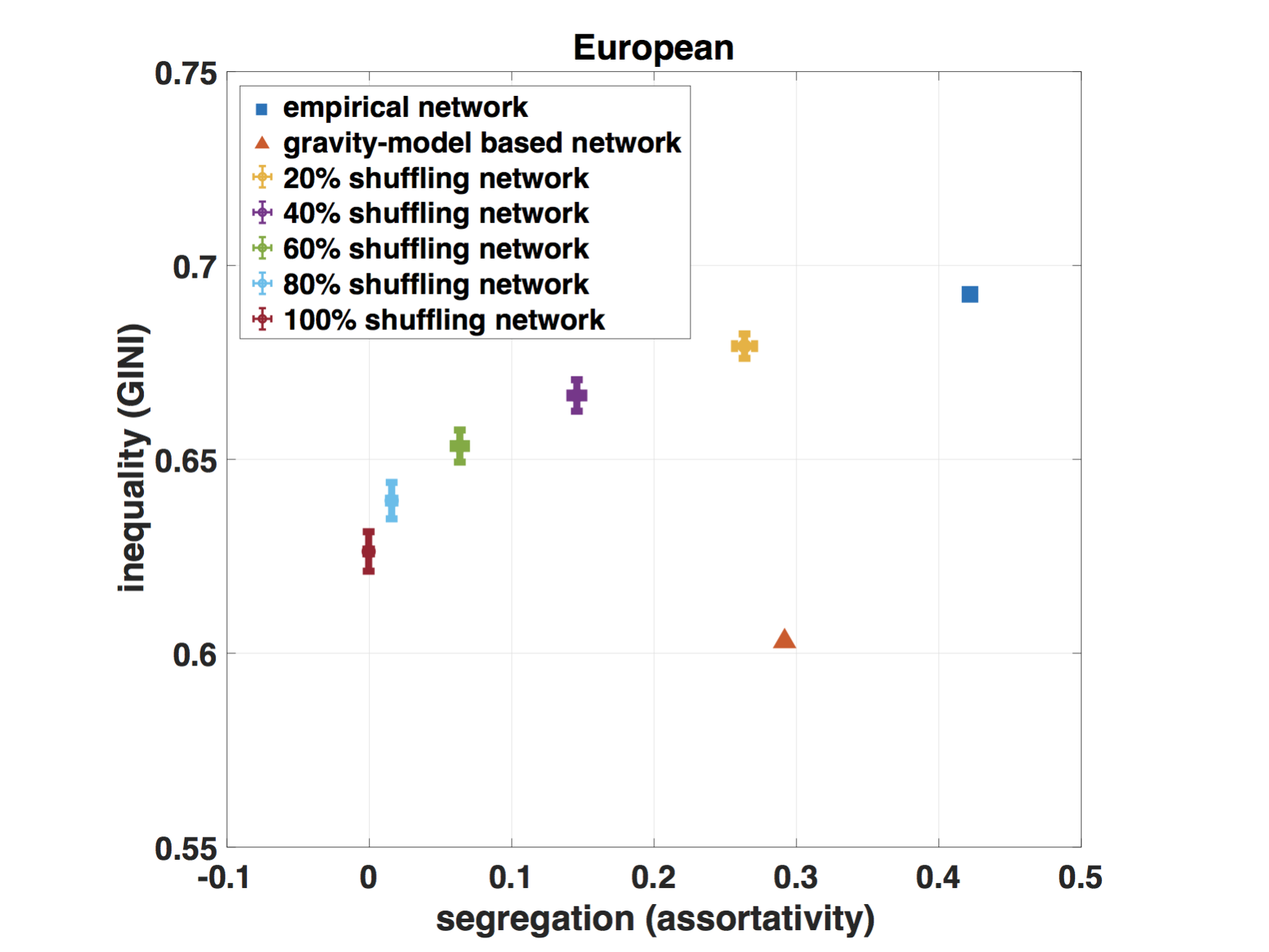}}
        \caption{Association of segregation (assortativity) and inequality (GINI coefficient) between neighborhoods in terms of total sales income in the European metropolitan area. {The error bars correspond to the standard deviations.}}
        \label{fig:figS16}
\end{figure}

\begin{figure}
      \centering
      {\includegraphics[width=16cm]{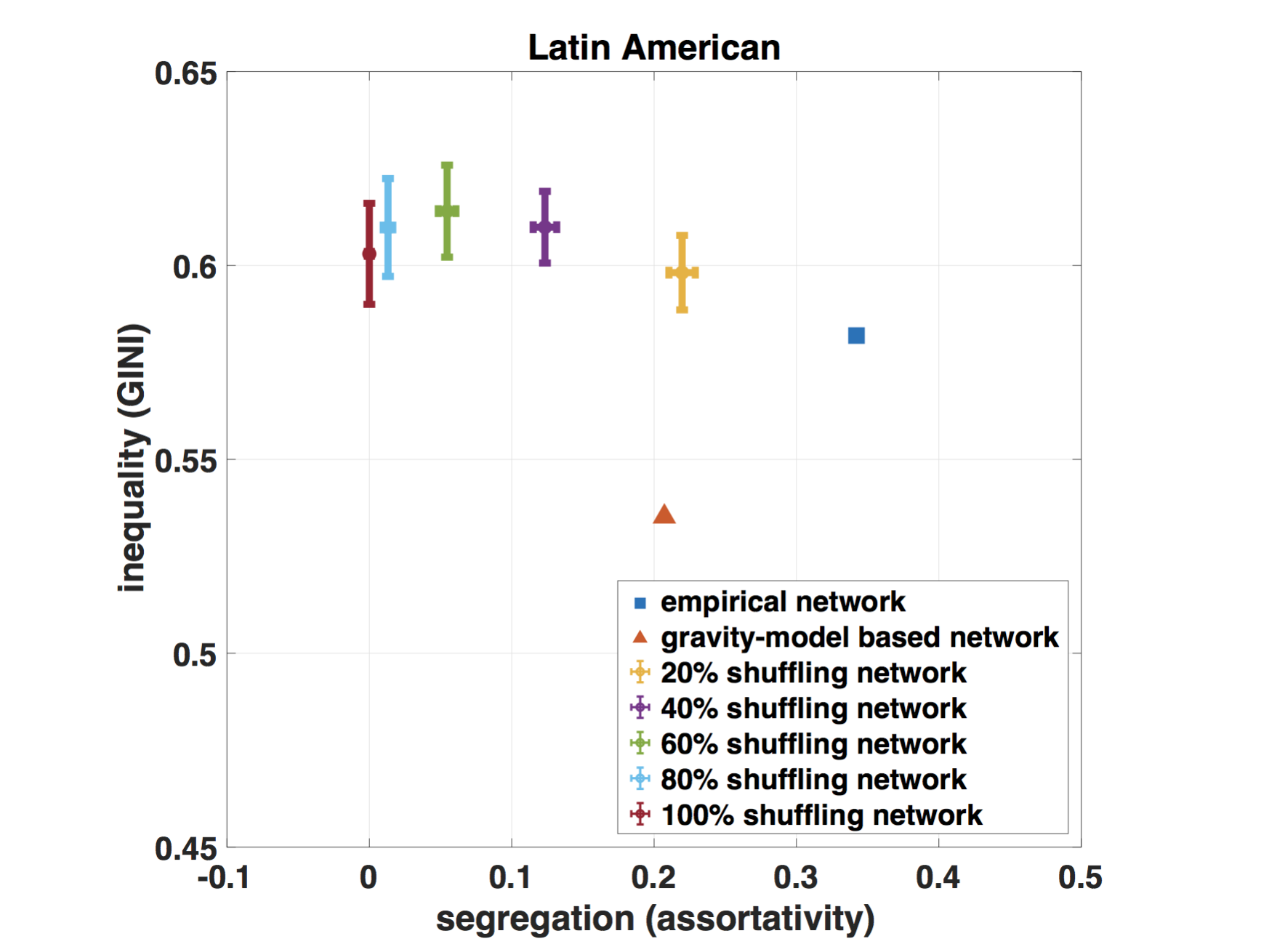}}
        \caption{Association of segregation (assortativity) and inequality (GINI coefficient) between neighborhoods in terms of total sales income in the Latin American metropolitan area. {The error bars correspond to the standard deviations.}}
        \label{fig:figS17}
\end{figure}

In Figure~\ref{fig:figS16} and Figure~\ref{fig:figS17} we present the GINI coefficient as a function of the assortativity for the empirical purchase network, the simulated purchase network based on the gravity model, and five networks based on random reshuffling of the customer and store locations.
{As we can see, in the European case, both inequality and segregation are highest in the empirical case, while in the gravity-model case they drop about 13\% and 31\%, respectively.} As we reshuffle the location of stores and customers, both the GINI coefficient and the assortativity decrease. However, while segregation goes all the way down to zero, a considerable degree of inequality persists. This is due to the inhomogeneous number of stores across neighborhoods.
{Similarly, in the Latin American case, inequality and segregation in the gravity-model case drop about 8\% and 40\%, respectively, comparing to the empirical case. However, while segregation goes down to zero, the GINI coefficient for the shuffling networks remain similar to the empirical case. This could be due to an even larger inhomogeneity in the number of stores across neighborhoods.
The results in the European case (and partially in the Latin American case) suggests that there might be a relationship between segregation pattern in purchase behavior and the level of inequality between neighborhoods in terms of their sales revenue.}

\clearpage
\beginsupplement

\clearpage

\clearpage

\clearpage

\clearpage

\clearpage

\clearpage

\clearpage

\clearpage

\clearpage

\clearpage

\begin{table}[h]
\caption{Model parameters obtained by an OLS fitting, for (Top) European metropolitan area, (Middle) Latin American metropolitan area, and (Bottom) Northern American metropolitan area.}
\centering

\begin{tabular}{c c | c c}
\hline
\multicolumn{4}{c}{European} \\
\hline
\multicolumn{2}{c|}{Credit card data set} & \multicolumn{2}{c}{Twitter data set} \\
\hline
$c_p$ & 0.249 & $c_t$ & 0.119 \\[0.1cm]
$\beta_{p1}$ & 0.762 & $\beta_{t1}$ & 0.594 \\[0.1cm]
$\beta_{p2}$ & 0.598 & $\beta_{t2}$ & 0.541 \\[0.1cm]
$\epsilon_p$ & 0.233 & $\epsilon_t$ & 0.029 \\[0.1cm]
$\alpha_p$ & 0.918 & $\alpha_t$ & 0.582 \\[0.1cm]
\hline\\
\end{tabular}

\begin{tabular}{c c | c c}
\hline
\multicolumn{4}{c}{Latin American} \\
\hline
\multicolumn{2}{c|}{Credit card data set} & \multicolumn{2}{c}{Twitter data set} \\
\hline
$c_p$ & 0.231 & $c_t$ & 0.085 \\[0.1cm]
$\beta_{p1}$ & 0.681 & $\beta_{t1}$ & 0.493 \\[0.1cm]
$\beta_{p2}$ & 0.824 & $\beta_{t2}$ & 0.829 \\[0.1cm]
$\epsilon_p$ & 1.026 & $\epsilon_t$ & 0.298 \\[0.1cm]
$\alpha_p$ & 1.058 & $\alpha_t$ & 0.633 \\[0.1cm]
\hline\\
\end{tabular}

\begin{tabular}{c c}
\hline
\multicolumn{2}{c}{Northern American} \\
\hline
\multicolumn{2}{c}{Twitter data set} \\
\hline
$c_t$ & 7.941 \\[0.1cm]
$\beta_{t1}$ & 0.276 \\[0.1cm]
$\beta_{t2}$ & 0.353 \\[0.1cm]
$\epsilon_t$ & 0.330 \\[0.1cm]
$\alpha_t$ & 0.837 \\[0.1cm]
\hline
\end{tabular}

\label{tab:params}
\end{table}

\end{document}